\documentclass[pra,twocolumn,superscriptaddress,longbibliography]{revtex4-2}
\usepackage[colorlinks=true, citecolor=blue, urlcolor=blue, linkcolor=blue ]{hyperref}
\usepackage{graphicx,amsmath,amssymb,bm,amsthm,mathrsfs,orcidlink}

\begin{document}
\title{Protecting Spin Squeezing from Decoherence}
\author{Lin Jiao}
\affiliation{Department of Physics and Astronomy, and Smalley-Curl Institute, Rice University, Houston, TX 77251-1892, USA}
\author{Han Pu}
\email{hpu@rice.edu}
\affiliation{Department of Physics and Astronomy, and Smalley-Curl Institute, Rice University, Houston, TX 77251-1892, USA}
\author{Jun-Hong An\orcidlink{0000-0002-3475-0729}}
\email{anjhong@lzu.edu.cn}
\affiliation{Key Laboratory of Quantum Theory and Applications of MoE, Lanzhou Center for Theoretical Physics, Key Laboratory of Theoretical Physics of Gansu Province, Gansu Provincial Research Center for Basic Disciplines of Quantum Physics, Lanzhou University, Lanzhou 730000, China}

\begin{abstract}
As a crucial resource in the field of quantum metrology, spin squeezing can facilitate highly precise measurements that surpass the limitations imposed by classical physics. However, the quantum advantage of spin squeezing is significantly compromised by decoherence, thus impeding its practical implementation. Here, by investigating the influence of local dissipative environment on spin squeezing beyond the conventional Born-Markov approximation, we find a mechanism to protect spin squeezing from decoherence and show that robust spin squeezing can be achieved in the steady state. We outline an experimental proposal to verify our prediction in a trapped-ion platform. Overcoming the challenges set by decoherence in spin squeezing, our work provides guidance to realize high-precision sensing in realistic environments and sheds new light on the effect of non-Markovian environment on quantum systems.
\end{abstract}
\maketitle

\section{Introduction}
The goal of quantum metrology is to achieve higher precision in measurements than is classically feasible by exploiting quantum resources. It is expected to lead to a number of significant technological breakthroughs \cite{RevModPhys.89.035002,RevModPhys.90.035005}. One such distinguished quantum resource is spin squeezing, which is closely related to many-body entanglement \cite{PhysRevLett.112.155304}. As a consequence of quantum correlations between particles, spin-squeezed states exhibit reduced quantum fluctuations of collective spins \cite{PhysRevA.46.R6797,PhysRevA.47.5138,PhysRevA.50.67,MA201189,PhysRevLett.112.155304} in comparison to spin coherent states. It holds promise for the development of revolutionary technologies, such as quantum gyroscope and gravimetry \cite{Rosi2014,10.1116/1.5120348,PhysRevLett.120.033601,PhysRevLett.125.100402}, atomic clocks \cite{PhysRevLett.112.103601,PhysRevLett.125.210503,Schulte2020,Malia2022,Eckner2023,Robinson2024}, and magnetometers \cite{PhysRevLett.113.103004,PhysRevX.5.031010,Bao2020,PhysRevLett.127.193601}. Recent experiments have realized spin squeezing in atomic ensembles comprising up to $10^{13}$ atoms \cite{Bao2020,Jin:21,Kong2020}. 

As in most quantum resources, the practical application of spin squeezing in large-scale quantum metrology is challenged by decoherence caused by various kinds of noise in the quantum world. Spin squeezing tends to deteriorate or even completely vanish under the influence of decoherence \cite{PhysRevA.81.022106,PhysRevA.86.023812,PhysRevResearch.5.043279,https://doi.org/10.1002/qute.202300218}, limiting both the stability \cite{PRXQuantum.4.020322} and scalability of the associated quantum metrology schemes \cite{PhysRevLett.127.160501,Bornet2023,PhysRevLett.133.210401}. Therefore, the pursuit of efficient strategies to protect spin squeezing from decoherence is of crucial importance. Active ways, including dynamical decoupling \cite{PhysRevA.86.012311}, weak measurement \cite{Liao_2017}, and reservoir engineering \cite{PhysRevLett.127.083602}, have been proposed to overcome the destructive effects of decoherence on spin squeezing. A common character of these works is that the description of decoherence was based on the Born-Markov approximation. Given the inherent non-Markovian nature of the decoherence dynamics \cite{PhysRevA.84.012103,PhysRevLett.115.170801,Rivas_2014,RevModPhys.88.021002,RevModPhys.89.015001,LI20181}, however, such a treatment may not be sufficient. In fact, the non-Markovian effect was found to play a positive role in preserving the quantum advantages of optical squeezing \cite{PhysRevLett.123.040402} and GHZ-type entanglement \cite{PhysRevLett.109.233601,PhysRevLett.129.070502,PhysRevLett.131.050801} in noisy quantum metrology. It is therefore necessary to investigate the effects of non-Markonivan environments on spin squeezing. 

In this paper, by investigating the non-Markovian decoherence dynamics of an atomic ensemble prepared in spin squeezed states, we discover a mechanism to protect the spin squeezing from decoherence. It is found that the robustness of the spin squeezing is essentially determined by the feature of the energy spectrum of the total system formed by each atom and its local environment. Contrary to conventional wisdom, our findings reveal that, while stronger coupling is typically detrimental in Markovian settings, it may help to preserve spin squeezing in structured non-Markovian environments. We propose testing this prediction in a trapped-ion platform. Our work provides a universal mechanism for realizing stable and scalable spin squeezing, crucial for high-precision quantum metrology in realistic noisy environments.

This paper is organized as follows. In Sec. \ref{s2}, we introduce the spin-squeezing parameter and briefly review the one- and two-axis twisted states. In Sec. \ref{s3}, we analyze the effects of dissipative environments beyond the Born-Markov approximation and demonstrate the bound-state mechanism that protects spin squeezing in the steady-state limit. In Sec. \ref{s4}, we discuss a possible implementation in a trapped-ion platform. Section \ref{s5} summarizes our main findings and conclusions. Additional technical details are provided in the Appendices.

\section{Spin squeezing}\label{s2}
Spin squeezed states are entangled states of collective spins whose variance in certain components perpendicular to the mean spin direction is smaller than that of a spin coherent state. Consider that an ensemble of $N$ two-level atoms or spin-$\tfrac{1}{2}$ particles described by the Pauli matrices $\hat{\pmb\sigma}_l=(\hat{\sigma}_l^x,\hat{\sigma}_l^y,\hat{\sigma}_l^z)$, with $l=1,\cdots, N$, is in a state $\rho$. Its mean spin direction is ${\bf n}_0=\langle\hat{\bf J}\rangle/|\langle\hat{\bf J}\rangle|$, where $\langle\hat{\bf J}\rangle=\text{Tr}(\rho \hat{\bf J})$ and $\hat{\bf J}=\sum_{l=1}^N\tfrac{\hat{\pmb\sigma}_l}{2}$ is the collective spin. The use of this state in Ramsey spectroscopy to measure a phase $\phi$ results in an error $\delta\phi=\min_\beta \Delta J_{\perp,\beta}/|\langle\hat{\bf J}\rangle|$ \cite{PhysRevA.46.R6797}, where $\Delta J_{\perp,\beta}=(\langle \hat{J}_{\perp,\beta}^2\rangle-\langle \hat{J}_{\perp,\beta}\rangle^2)^\frac{1}{2}$ is the variance of $\hat{J}_{\perp,\beta}=(\cos\beta {\bf n}_1+\sin\beta {\bf n}_2)\cdot \hat{\bf J}$, with ${\bf n}_{1,2}$ being two orthogonal unit vectors perpendicular to ${\bf n}_0$. For a spin coherent state $|\theta,\varphi\rangle={e^{\zeta \hat{J}_-}\over (1+|\zeta|^2)^{j}}|j,j\rangle$, with $\zeta=e^{i\varphi}\tan{\theta\over 2}$, $\hat{J}_\pm=\hat{J}_x\pm i\hat{J}_y$, $j=\tfrac{N}{2}$, and $|j,j\rangle$ being the common eigenstate of $\hat{J}^2$ and $\hat{J}_z$, the phase error is $\delta\phi_\text{SCS}=N^{-1\over2}$, which is the shot-noise limit (SNL). Wineland \textit{et al.} defined the ratio between $\delta\phi$ and $\delta\phi_\text{SCS}$, i.e., $\xi \equiv \sqrt{N} \min_\beta\Delta {J}_{\perp,\beta}\big/|\langle\hat{\bf J}\rangle|$, as the spin squeezing parameter \cite{PhysRevA.50.67}. If $\xi<1$, the state is spin squeezed and the phase error $\delta \phi = \xi/\sqrt{N}$ exceeds the SNL.

Two types of spin squeezed states, i.e., one-axis twisted (OAT) and two-axis twisted (TAT) states, are widely studied \cite{PhysRevA.47.3554,PhysRevA.40.2417,PhysRevA.92.033826,PhysRevA.63.042304,PhysRevLett.101.073601,PhysRevLett.99.163002,PhysRevA.79.021603}. They have a mean spin direction along the $z$-axis and are generated from $|j,-j \rangle$ as
\begin{eqnarray}
   |\Psi_{\text{OAT}}\rangle &=& e^{-i\Theta \hat{J}_{x}^{2}} \,|j,-j\rangle \,, \label{psioat} \\
   |\Psi_{\text{TAT}}\rangle &=& e^{\Theta(\hat{J}_{+}^{2}-\hat{J}_{-}^{2})}\,|j,-j\rangle \,. \label{psitat}
\end{eqnarray}
The OAT state has been realized in systems of atoms \cite{PhysRevLett.101.073601,PhysRevLett.99.163002,doi:10.1126/science.1058149}, trapped ions \cite{doi:10.1126/science.aad9958,Lu2019,Figgatt2019}, and superconducting qubits \cite{doi:10.1126/science.aay0600,doi:10.1126/sciadv.aba4935}. We can calculate $\langle\hat{\bf J}\rangle=-j\cos^{2j-1}\Theta(0,0,1)$ and $\min_\beta\Delta J_{\perp,\beta}^2=(\langle\hat{J}_x^2+\hat{J}_y^2\rangle-|\langle\hat{J}_-^2\rangle|)/2$. The permutation symmetry of the $N$ identical spins causes
\begin{eqnarray}
\langle\hat{J}_\alpha^2\rangle&=&N[1+(N-1)\langle \hat{\sigma}_{1}^\alpha\hat{\sigma}_{2}^\alpha\rangle]/4, \\
\langle\hat{J}_-^2\rangle&=&N(N-1)\langle \hat{\sigma}_{1-}\hat{\sigma}_{2-}\rangle,\label{tpmn}
\end{eqnarray}
with $\hat{\sigma}_{l-} \equiv(\hat{\sigma}_l^x-i\hat{\sigma}_l^y)/2$. One easily obtains $\langle \hat{\sigma}_{1-}\hat{\sigma}_{2+}\rangle=A/8$ and $\langle \hat{\sigma}_{1-}\hat{\sigma}_{2-}\rangle=(iB-A)/8$, with $A\equiv 1-\cos^{N-2}(2\Theta)$ and $B \equiv 4\sin\Theta\cos^{N-2}\Theta$. It follows
\begin{eqnarray}
&&\xi_{\rm OAT}^2\simeq1+(N-1)(A-\sqrt{A^2+B^2})/4.\label{xir1}
\end{eqnarray}
Under the condition $N\Theta^2\ll 1 \ll N\Theta$, it tends to $\xi_{\rm OAT}^2\simeq(N\Theta)^{-2}+N^2\Theta^4/6$. Its minimum is $\min\xi_{\rm OAT}^2\simeq 1.04 N^{-2\over3}$ when $\Theta=\Theta_0 \equiv 3^{1/6}/N^{2/3}$. Thus, we achieve $\delta\phi_\text{OAT}\propto N^{-5/6}$ using $|\Psi_{\text{OAT}}\rangle$ in Ramsey spectroscopy.

The TAT state is less analytically tractable. The numerical fitting shows that its spin squeezing parameter scales with the atom number as $\xi_{\rm TAT}\simeq \sqrt{2}N^{-1/2}$ in the large-$N$ limit. Thus, its phase error reaches the Heisenberg limit $\delta\phi_\text{TAT}\simeq \sqrt{2}N^{-1}$. Several schemes have been proposed to transform the one-axis twisting into a two-axis one \cite{PhysRevLett.107.013601,PhysRevA.90.013604,PhysRevA.91.043642,PhysRevA.107.042613,Xu2017}. $|\Psi_\text{TAT}\rangle$ has been generated in an ensemble of 700 rubidium atoms \cite{Luo2025}.

\section{Effects of environment}\label{s3}
The advantages of spin squeezing in quantum metrology are challenged by the ubiquitous decoherence in the quantum world. It has been found that the system-environment interplay caused by the inherent non-Markovian nature can induce diverse characters absent in the Born-Markov approximation \cite{PhysRevA.81.052330,PhysRevLett.109.170402,PhysRevLett.121.220403}. To reveal the practical performance of spin squeezing in Ramsey spectroscopy, we go beyond the widely used Born-Markov approximation and investigate the impact of dissipative decoherence on spin squeezing.

We consider that each of $N$ spins initially in $|\Psi_{\text{OAT}}\rangle$ or $|\Psi_{\text{TAT}}\rangle$ is coupled to a dissipative environment. We ignore the decoherence during the state preparation due to its very short time duration \cite{PhysRevA.108.062611,Schulte2020ramsey}. Note that even if decoherence affects state preparation so that the initial state deviates away from $|\Psi_{\text{OAT}}\rangle$ or $|\Psi_{\text{TAT}}\rangle$, the physics discussed below would not change. The Hamiltonian of the global system is $\hat{H}=\sum_{l=1}^N\hat{H}_l$ with
\begin{equation}
\hat{H}_l=\omega_{0}\hat{\sigma}_{l+}\hat{\sigma}_{l-}+\sum_k[\omega_{lk}\hat{a}_{lk}^{\dagger}\hat{a}_{lk}+(g_{lk}\hat{a}_{lk}\hat{\sigma}_{l+}+\text{h.c.})],\label{totalH}
\end{equation}
is the total Hamiltonian for the $l$th spin and its environment. Here, $\hat{a}_{lk}$ is the annihilation operator of the $k$th mode with the frequency $\omega_{lk}$ of the environment felt by the $l$th spin. The environmental spectral density is defined as $J_l(\omega)=\sum_k|g_{lk}|^2\delta(\omega-\omega_{lk})$. Inspired by the realization of the Ohmic spectral density in trapped ion \cite{Sun2025} and circuit QED \cite{FornDiaz2017,Magazzù2018} systems, we assume that the environments possess a common Ohmic-family spectral density $ J_l(\omega)\equiv J(\omega)=\eta\omega^{s}\omega_{c}^{1-s} e^{-\omega/\omega_{c}}$, where $\eta$ is a dimensionless coupling constant, $\omega_{c}$ a cutoff frequency, and $s$ an Ohmicity index \cite{RevModPhys.59.1}. 

When the environments are initially in the vacuum state $|\{0_k\}\rangle$, the calculation is simplified. If the spin (we omit ``$l$'' for brevity) is initially in $|\downarrow \rangle$, then the state remains in it as it is an eigenstate of Eq. \eqref{totalH}. If the spin is initially in $|\uparrow \rangle$, then the total state evolves as $|\Psi(t) \rangle = u(t) |\uparrow,  \{0_k\} \rangle+ \sum_k v_k(t)|\downarrow,1_k \rangle$ under $u(0)=1$. Combining the two cases, we can trace the environmental degrees of freedom for any initial spin state and obtain a non-Markovian master equation, see Appendix \ref{sec2},
\begin{equation}
\dot{\rho}(t)=\sum_{l=1}^N\{-i\Omega(t)[\hat{\sigma}_{l+}\hat{\sigma}_{l-},\rho(t)]+\Gamma(t)\check{\mathcal L}_l\rho(t)\},\label{nmeqat}
\end{equation}
where $\check{\mathcal L}_l\cdot=2\hat{\sigma}_{l-}\cdot\hat{\sigma}_{l+}-\{\hat{\sigma}_{l+}\hat{\sigma}_{l-},\cdot\}$ is the Lindblad superoperator and $\Gamma(t)+i\Omega(t)=-\dot{u}(t)/u(t)$. As the time-dependent factor of quantum coherence, $u(t)$ satisfies
\begin{equation}
\dot{u}(t)+i\omega_0u(t)+\int_{0}^{t}f(t-\tau)u(\tau)d\tau=0,  \label{uleq}
\end{equation}
with $f(t-\tau)=\int_{0}^{\infty}J(\omega)e^{-i\omega (t-\tau)}d\omega$ being the environmental correlation function. The non-Markovian effect has been incorporated in the time-dependent coefficients of Eq. \eqref{nmeqat}. The Kraus representation renders the solution of Eq. \eqref{nmeqat} as $\rho(t)=\check{\Lambda}_t^{\otimes N}\rho(0)$, where $\check{\Lambda}_t\cdot=\sum_{\mu=1}^2\hat{K}_\mu\cdot\hat{K}^\dag_\mu$ with $ \hat{K}_{1}=\text{diag}[u(t), 1]$ and $\hat{K}_{2}=[1-|u(t)|^2]^{1/2}\hat{\sigma}_-$ \cite{Kraus1983StatesEA}. Because the permutation symmetry is preserved in the presence of environments, we can still evaluate the expectation values of the quadratic forms of collective spins by those of the bipartite operators $\langle \hat{\sigma}_{1}^\alpha\hat{\sigma}_{2}^\alpha\rangle=\text{Tr}[ \hat{\sigma}_{1}^\alpha\hat{\sigma}_{2}^\alpha\check{\Lambda}_t^{\otimes 2}\rho(0)]=\langle(\check{\Lambda}_t^{\dag}\hat{\sigma}^{\alpha})^{\otimes 2}\rangle_0$, where $\langle\cdot\rangle_0=\text{Tr}[\cdot\rho(0)]$. 

Consider first the initial state to be $|\Psi_{\rm OAT} \rangle$, i.e., $\rho(0) =|\Psi_{\rm OAT} \rangle \langle \Psi_{\rm OAT}|$. Using $\check{\Lambda}_t^\dag\hat{\sigma}_-=u(t)\hat{\sigma}_-$, Appendix \ref{sec4} gives
\begin{eqnarray}\label{eq16}
\xi_{\rm OAT}^2(t)=1+|u(t)|^2(N-1)(A-\sqrt{A^2+B^2})/4\,.
\end{eqnarray}
When the system-environment coupling is weak and the correlation time in $f(t-\tau)$ is much shorter than the typical time scale of the spins, we can apply the Born-Markov approximation \cite{PhysRevLett.102.040403,PhysRevLett.108.130402}, under which the solution of Eq.~\eqref{uleq} reads $u_\text{BMA}(t)=e^{-[\kappa+i(\omega_0+\Delta(\omega_0))]t}$ with $\kappa=\pi J(\omega_0)$, $\Delta(\omega_0)=\mathcal{P}\int_{0}^{\infty}\frac{J(\omega)}{\omega_0-\omega}d\omega$, and $\mathcal{P}$ being the Cauchy principal value. Substituting this into Eq.~\eqref{eq16}, we obtain that $\xi_{\rm OAT}^2(t)$ exponentially increases to one with a rate $2\kappa \propto \eta$. The spin squeezing is destroyed with time. The stronger the coupling strength is, the faster it is destroyed. Physically, this means the initial quantum coherence in $\rho(0)$ is irreversibly lost to the environment. Hence, it is natural to expect that the performance of quantum sensing in practical noisy settings is hampered by decoherence.  The result matches the general belief of decoherence effects on quantum sensing schemes \cite{PhysRevA.95.053837,PhysRevLett.108.130402,PhysRevA.90.033846}. 

Does this result hold in general? To address this, we go beyond the above approximation. In this case, Eq.~\eqref{uleq} can only be solved numerically. However, its asymptotic form is solvable by the Laplace transform, which converts Eq.~\eqref{uleq} into $\tilde{u}(z)=[z+i\omega_0+\int_{0}^{\infty}\frac{J(\omega)d\omega}{z+i\omega}d\omega]^{-1}$. Then $u(t)$ is obtained by performing the inverse Laplace transform to $\tilde{u}(z)$. This requires finding the poles of $\tilde{u}(z)$, which, by setting $E=iz$, is determined by the equation 
\begin{equation}
    Y(E)=E,\label{ddeig}
\end{equation}with $Y(E)\equiv \omega_0-\int_{0}^{\infty}\frac{J(\omega)}{\omega-E}d\omega$. It is interesting to find that Eq. \eqref{ddeig} is the same as the equation satisfied by the eigenenergy of Eq.~\eqref{totalH}. Explicitly, the eigenstate of the $l$th subsystem in the single-excitation subspace can be expended as 
\begin{equation}
   |\Phi\rangle=(c\hat{\sigma}_{l+}+\sum_k d_{k}\hat{a}^\dag_{lk})\left|\downarrow_l,\{0_k\}_l\right\rangle ,\label{ddeigt}
\end{equation}where $c$ and $d_k$ are the probability amplitudes to be determined. Inserting it into $\hat{H}_l|\Phi\rangle=E|\Phi\rangle$, we obtain $Ec= \omega_{0}c+\sum_{k}g_{lk}d_{k}$ and $Ed_{k} = \omega_{k}d_{k}+g_{lk}^{*}c$. These equations easily lead to $E =\omega_0-\sum_k \frac{|g_{lk}|^2}{\omega_k-E}$, which in the continuous limit of the environmental frequencies is just Eq. \eqref{ddeig}. Note that $Y(E)$ is ill-defined and oscillates rapidly between $\pm\infty$ in the regime $E >0$ due to the divergence of the integrand in $Y(E)$. Thus, we obtain infinite roots $E$ in this regime, which form a continuum energy band. In the regime $E<0$, $Y(E)$ is a monotonic decreasing function of $E$. It has an isolated root $E_b$ provided $Y(0)<0$, which represents a bound state. Substituting these poles into the inverse Laplace transform, we obtain
\begin{equation}
u(t)=Ze^{-iE_{b}t}+\int_{0}^{\infty}\tfrac{J(E)e^{-iEt}}{[E-\omega_0-\Delta(E)]^{2}+[\pi J(E)]^{2}}dE,\label{ut}
\end{equation} 
where $Z=[1+\int_{0}^{\infty}\frac{J(\omega)d\omega}{(E_{b}-\omega)^2}]^{-1}$ originates from the bound state. The second term is from the continuum energy and tends to zero in the long-time limit due to out-of-phase interference. Thus, we have 
\begin{equation}  \lim_{t\rightarrow\infty}u(t)= \left\{\begin{array}{ll} 0\,; & {\rm without \;\,bound \;\,state} \\  Ze^{-iE_{b}t}\,; & {\rm with \;\,bound \;\,state} \end{array} \right.  .\label{uttt}
\end{equation}
For the Ohmic-family spectral density, the condition to form the bound state, i.e., $Y(0)<0$, amounts to 
\begin{equation}
    \eta > \eta_c \equiv \omega_0/[\omega_c\gamma(s)] \,, \label{etac}
\end{equation}
with $\gamma(s)$ being the Gamma function. The result demonstrates that the dynamics of the open system characterized by $u(t)$ is intrinsically determined by the feature of energy spectrum of the total system. Actually, this is generally valid for open systems, including the two-level system and the harmonic oscillator \cite{PhysRevLett.133.050401}, coupled to the environment that conserves the total excitation number regardless of the form of the spectral density. The derivation is given in Appendix \ref{sec2}. 

\begin{figure}[tbp]
\centering
\includegraphics[height=.55\columnwidth]{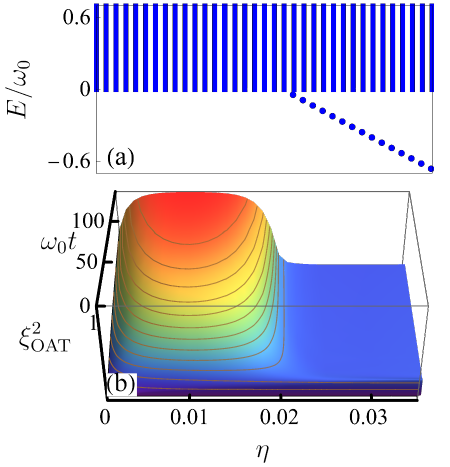}\hspace{-0.1cm}\includegraphics[height=.55\columnwidth]{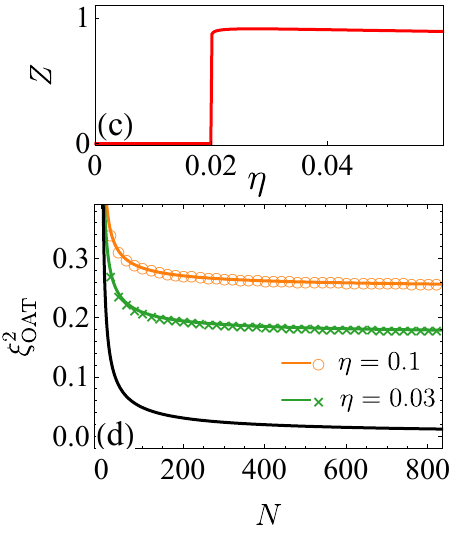}\\
\includegraphics[width=\columnwidth]{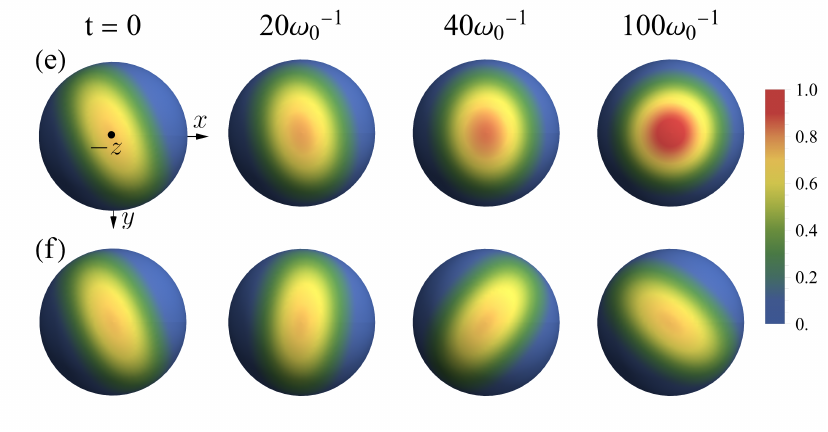}\\
\includegraphics[height=0.45\columnwidth]{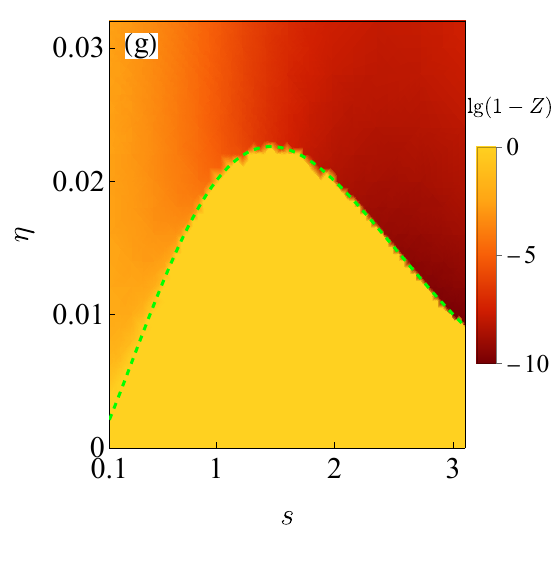}~
\includegraphics[height=0.45\columnwidth]{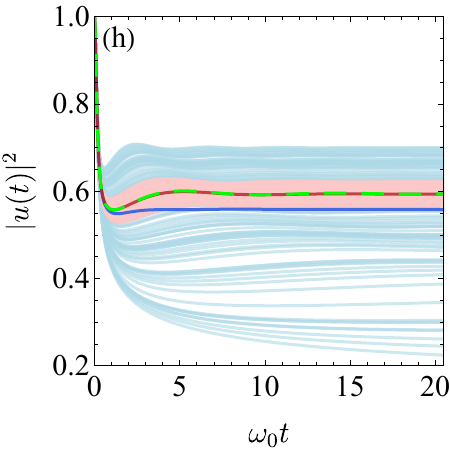}~
\caption{(a) Energy spectrum, (b) evolution of $\xi_{\rm OAT}^2(t)$ for $N=100$, and (c) $Z$ as a function of $\eta$. (d) $\xi_{\rm OAT}^2(\infty)$ at $t=400\omega_0^{-1}$ as a function of $N$ for different $\eta$. Symbols are numerical results, colored lines are analytic results from Eq.~\eqref{minxi}, and black line is the decoherence-free result. Normalized $Q(\theta,\varphi)$ at different time when $\eta=0.01$ (e) and $0.03$ (f) for $N=10$. We use $\omega_c=50\omega_0$ and $s=1$, which gives $\eta_c=0.02$, and $\Theta=\Theta_0$. (g) $\log_{10}(1-Z)$ as a function of $\eta$ and $s$ with $\omega_c=50\omega_c$. The green dotted line is $\eta_c$. (h) Evolution of $|u(t)|^2$ when $\omega_c=10\omega_0$, $s=1$, and $\eta=0.2$ in the presence of the dephasing noise. The pink and light-blue regions are the 100 stochastic realizations for $w=0.2\omega_0$ and $\omega_0$ with their averages denoted by the red dashed and blue solid lines, respectively. The green line is the dephasing-free result. }\label{p2}
\end{figure}

Equations (\ref{uttt}) and~\eqref{eq16} exhibit the essential role of the bound state. When it is absent, $\xi_{\rm OAT}$ approaches one, like the Born-Markov result. When it is present, $\xi_{\rm OAT}^2(\infty) \simeq 1+Z^{2}(N^{-2}\Theta^{-2}+\frac{N^{2}\Theta^{4}}{6}-1)$ has a minimum
\begin{equation}
\xi_{\rm OAT}^2(\infty)|_{\Theta=\Theta_0}\simeq 1.04Z^2N^{-{2\over 3}}+1-Z^2.\label{minxi}
\end{equation}
Hence, it is desirable to make $Z$ as close to 1 as possible by optimizing the parameters in the spectral density via reservoir engineering. $\xi_{\rm OAT}^2(\infty)<1$ indicates that spin squeezing survives in the steady state. Physically, this is due to the fact that the bound state partially preserves the quantum coherence in the initial state.   

We make numerical calculations by choosing an Ohmic spectral density. Qualitatively similar results are obtained for other spectra. The energy spectrum of each spin and its environment in Fig.~\ref{p2}(a) indicates that a bound state is formed when $\eta > \eta_c$. By numerically solving Eq.~\eqref{uleq}, the dynamics in Fig.~\ref{p2}(b) confirms that $\xi^2_{\rm OAT}(t)$ approaches one in the absence of the bound state and stabilizes at a $\eta$-dependent value less than one in the presence of the bound state. We also see from Fig.~\ref{p2}(c) that $Z$ exhibits a sudden jump from zero to a finite value when $\eta>\eta_c$. Choosing a sufficiently large $t$, we plot $\xi_{\rm OAT}^2(\infty)$ as a function of the spin number $N$ in Fig.~\ref{p2}(d). The numerical result is in perfect agreement with Eq.~\eqref{minxi}. To visually depict the evolution of the spin squeezing, we calculate the Husimi Q function $Q(\theta,\varphi)=\tfrac{2j+1}{4\pi}\langle\theta,\varphi|\rho(t)|\theta,\varphi\rangle$, which maps $\rho(t)$ to a quasiclassical probability distribution in the phase space defined by the spin coherent state $|\theta,\varphi\rangle$, see Appendix \ref{sec5}. For $\eta < \eta_c$, $Q(\theta,\varphi)$ tends to an isotropically distributed variance, which reflects vanishing of the spin squeezing [see Fig. \ref{p2}(e)]. For $\eta > \eta_c$, on the other hand, $Q(\theta,\varphi)$ preserves the spin squeezing in the steady state [see Fig.~\ref{p2}(f)]. The value of $Z$ can be maximized as close as possible to one by optimizing the parameters in the spectral density [see Fig.~\ref{p2}(g)]. Our numerical result confirms the crucial role played by the bound state in preventing spin squeezing from the destruction of dissipative noise.

In addition to the dissipation in Eq.~\eqref{totalH}, dephasing is another type of common source of decoherence \cite{PhysRevLett.109.233601}. A rigorous treatment of simultaneous dissipation and dephasing is challenging. Here we phenomenologically model the dephasing by adding a classical noise $w\chi\hat{\sigma}_l^z$ in Eq. \eqref{totalH}, where $\chi$ is a random number uniformly distributed in $[-1,1]$ and $w$ is the noise strength. This results in a dephasing time $T_2=w^{-1}$. The dephasing leads to a broadening of $|u(t)|^2$, but its average remains stable in the long-time limit even when the noise is as strong as $w=\omega_0$ [see Fig. \ref{p2}(h)]. It demonstrates the robustness of our bound state mechanism in protecting the spin squeezing against the dephasing. 

\begin{figure}[tbp]
\centering
\includegraphics[width=.5\columnwidth]{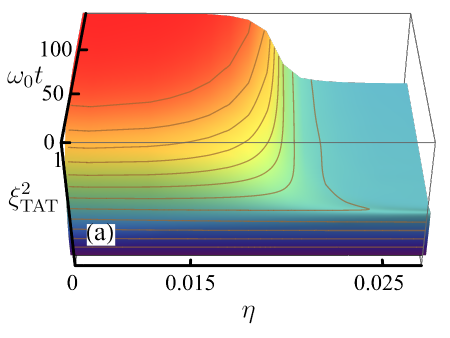}\hspace{-0.10cm}\includegraphics[width=.49\columnwidth]{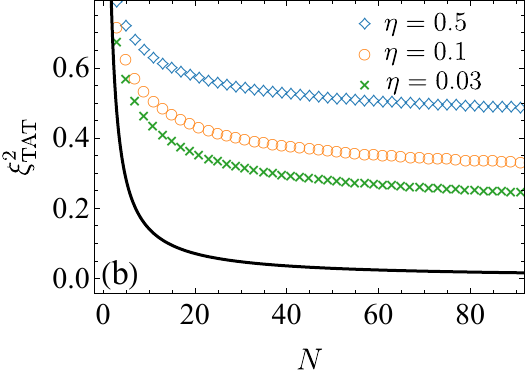}\\
\includegraphics[width=\columnwidth]{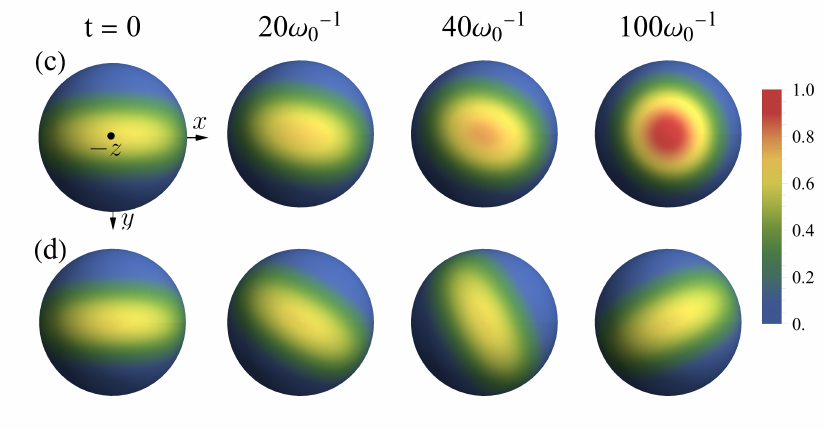}\\
\caption{(a) Evolution of $\xi_{\rm TAT}^2(t)$ for $N=50$. (b) $\xi_{\rm TAT}^2(\infty)$ at $t=400\omega_0^{-1}$ as a function of $N$. Normalized $Q(\theta,\varphi)$ at different times when $\eta=0.01$ (c) and $0.03$ (d) for $N=10$. Other parameters are the same as Fig. \ref{p2}.}\label{FigTAT}
\end{figure}

So far, we have discussed the environmental effects on $|\Psi_{\rm OAT} \rangle$. Similar studies can be carried out for $|\Psi_{\rm TAT} \rangle$. Using the Kraus representation, we find
\begin{equation}
\xi_{\rm TAT}^2(t)=1-|u(t)|^{2}+\tfrac{2|u(t)|^{2}}{N}(\langle \hat{J}_x^2+\hat{J}_y^2 \rangle_{0}-|\langle\hat{J}_{-}^{2}\rangle_{0}|), \label{tat}
\end{equation}
which is derived analytically in Appendix \ref{sec4}. Having the same energy spectrum as in the case of $|\Psi_\text{OAT}\rangle$, the bound state is present when $\eta>\eta_c$. Thus, $\xi^2_{\rm TAT}(t)$ also tends to one under the Born-Markov approximation and in the absence of the bound state. When the bound state is formed, $\xi^2_{\rm TAT}(t)$ tends to Eq.~\eqref{tat} with $u(t)$ replaced by $Z$. In the large-$N$ limit, $\xi^2_{\rm TAT}(\infty) =1-Z^2$, being the same as the OAT. Figures \ref{FigTAT}(a) and \ref{FigTAT}(b) show that the vanishing fate of spin squeezing is overturned by the formation of the bound state for $\eta>\eta_c$. The evolution of $Q(\theta,\varphi)$ in Figs. \ref{FigTAT}(c) and \ref{FigTAT}(d) confirms the dominant role of the bound state in protecting the spin squeezing from decoherence.

\begin{figure}[tbp]
\centering
\includegraphics[width=\columnwidth]{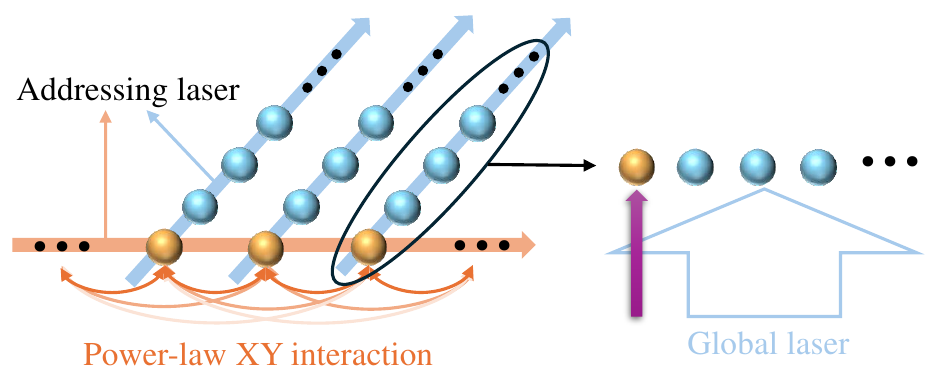}
\caption{An implementation of Eq. \eqref{totalH} on an trapped-ion platform. The edge ion chain (orange spheres) exhibits power-law XY interactions (orange lines) leading to the formation of a squeezed spin state. Each row of ions (blue spheres)  confined by beams (blue arrows) act as an environment. A narrow beam (magenta arrow) is used to control the coupling between each edge ion and its environment.}\label{scheme}
\end{figure}

\section{Experimental implementation}\label{s4} We now consider the possible experimental implementation.
The bound state and its dynamical effect have been observed in circuit QED systems \cite{nat.phys.1} and ultra-cold atom systems \cite{nature5,Kwon2022,Kim2025}. Spin squeezing in atomic ensembles has achieved levels of up to 20 dB \cite{Hosten2016}, with free evolution times that extend up to 1 second \cite{PRXQuantum.4.020322}. The Ohmic spectral density can be effectively approximated by a sum of Lorentzian lines determined by the motional modes of the structured environment \cite{PhysRevLett.120.030402,Lemmer2018} and has been simulated in circuit QED \cite{FornDiaz2017,Magazzù2018} and trapped ion \cite{Sun2025} systems. Based on existing experimental evidence \cite{Franke2023,PhysRevA.109.062402,Bruzewicz2016,Sun2025}, we propose a feasible scheme utilizing a two-dimensional (2D) trapped-ion platform to verify our result, as sketched in Fig.~\ref{scheme}. The efficient loading of a small 2D ion array has been experimentally demonstrated \cite{Bruzewicz2016}. The array consists of $N$ 1D ion chains. The $N$ ions at the end of each chain, depicted as the orange spheres in Fig.~\ref{scheme}, form the system spins. They can be prepared in a spin squeezed state via a global addressing beam which induces power-law transverse-field Ising interactions between spins \cite{Franke2023}. The rest of the ions in each chain serve as the environment for the corresponding edge ion, and together they realize the Hamiltonian (\ref{totalH}) in each chain. A single chain governed by such a Hamiltonian was recently realized \cite{PhysRevA.109.062402,Sun2025}, where an environment with an Ohmic spectral density was engineered. The parameters are tuned so that the condition for forming the bound state is met. These experimental advances provide strong supports for the realizability of our result. 

\section{Discussion and conclusion}\label{s5} In summary, we have demonstrated a non-Markovian mechanism to protect spin squeezing from dissipative environments. Although we have modeled the environment with an Ohmic spectral density, this mechanism also works for other spectral densities. We found that the dynamics of the spin squeezing for both the one- and two-axis twisted models depends sensitively on the feature of the energy spectrum of the total spin-environment system. Counterintuitively, to preserve spin squeezing, one needs to increase the system-environment coupling strength over a critical value such that a bound state is formed in the composite system. This is in stark contrast to the conventional wisdom obtained from the Born-Markov approximation, which tells us that to preserve quantum coherence, one needs to reduce the coupling to the environment as much as possible. Note that previous studies have explored the non-Markovian effects of both individual \cite{XUE20131328,Li_2018} and correlated \cite{PhysRevA.86.012308,PhysRevA.101.022327} decoherence on spin squeezing, but these works have not noticed the importance of the bound state and hence none of them achieves spin squeezing in the long-time limit. Our mechanism breaks the constraint of decoherence in quantum metrology and provides a guidance for developing spin-squeezing-based quantum sensing technologies. More generally, we hope that our work will motivate further studies of the non-Markovian quantum engineering. 

Our analysis has deliberately focused on an exactly solvable model in which each spin couples to an identical, independent environment and the spin–environment interaction conserves the total excitation number. This setting isolates the role of bound-state formation and allows us to demonstrate unambiguously that non-Markovian dissipation can stabilize spin squeezing in the steady state. It is nevertheless an important open problem to understand how this mechanism is modified in more realistic situations. In particular, in trapped-ion and related platforms, one expects residual cross-talk between nominally local environments, or even partially shared reservoirs. Extending our analysis to correlated or common environments and clarifying whether bound-state–assisted protection of squeezing survives in that case is a natural next step. Realistic ensembles will also exhibit inhomogeneities in spin frequencies and environmental parameters. Thus, quantifying the robustness of the steady-state squeezing against such inhomogeneity is crucial for future metrological applications. Finite reservoir temperature is another practically important extension, as thermal absorption in addition to emission is expected to reduce the steady-state squeezing with increasing $T$. Finally, it will be highly desirable to develop concrete reservoir-engineering strategies that transform natural decohering environments into environments supporting bound states for many spins. We leave these questions to future work.

\section*{Acknowledgments}
We acknowledge Si-Yuan Bai for the insightful discussions and Mingjian Zhu for suggestions. HP acknowledges support from U.S. NSF and the Welch Foundation (Grant No. C-1669). JHA is supported by the National Natural Science Foundation of China (Grants No. 12275109, No. 92576202, and No. 12247101), the Quantum Science and Technology-National Science and Technology Major Project (Grant No. 2023ZD0300904), the Natural Science Foundation of Gansu Province (Grants No. 26RCKA011 and No. 22JR5RA389), and the Fundamental Research Funds for the Central Universities (Grant No. lzujbky-2024-jdzx06).

\appendix

\section{Dynamics}\label{sec2}
Firstly, we consider two special cases of the initial state of the $l$th subsystem. Each of the environment is prepared in the vacuum state $|\{0_k\}_l \rangle$ initially. The spin can be in its ground state $|\downarrow_l \rangle$ or in the excited state $|\uparrow_l \rangle$.

\begin{enumerate}
\item The initial state is $|\Psi _{1}(0)\rangle=|\downarrow_{l},\{0_{k}\}_{l}\rangle $. As an eigenstate of $\hat{H}_{l}$ with eigenenergy zero, it does not change with time. Thus, we have $|\Psi_{1}(t)\rangle =|\Psi _{1}(0)\rangle $.

\item The initial state is $|\Psi _{2}(0)\rangle=|\uparrow_{l},\{0_{k}\}_{l}\rangle $. Because the total exitation number $\hat{\mathcal N}=\hat{\sigma}_{l+}\hat{\sigma}_{l-}+\sum_k\hat{a}_{lk}^{\dag}\hat{a}_{lk}$ is conserved, its evolved state is expanded as
\begin{equation}
|\Psi _{2}(t)\rangle =u(t)|\uparrow_{l},\{0_{k}\}_{l}\rangle+\sum_{k}v_{k}(t)|\downarrow_{l},1_{k,l}\rangle ,\label{apptmp2}
\end{equation}%
where $|1_{k,l}\rangle$ denotes the state of $l$th environment with only one excitation in its $k$th mode. From the Schr\"{o}dinger equation $i|\dot{\Psi}(t)\rangle=\hat{H}_l|\Psi(t)\rangle$, we obtain 
\begin{eqnarray}
i\dot{u}(t) &=&\omega _{0}u(t)-\sum_{k}g_{k}v_{k}(t),  \label{xc} \\
i\dot{v}_{k}(t) &=&\omega _{k}v_{k}(t)-g_{lk}u(t).  \label{dd}
\end{eqnarray}
Substituting the solution of Eq. (\ref{dd}) under $v_k(0)=0$ as $v_{k}(t)=ig_{lk}\int_{0}^{t}d\tau e^{-i\omega _{k}(t-\tau )}u(\tau )$ into Eq. (\ref{xc}), we have
\begin{equation}
\dot{u}(t)+i\omega _{0}u(t)+\int_{0}^{t}d\tau f(t-\tau )u(\tau )=0,\label{cctd}
\end{equation}
under $u(0)=1$, where $f(t-\tau ) =\int_0^\infty d\omega J(\omega )e^{-i\omega (t-\tau )}$ and $J(\omega ) =\sum_{k}|g_{lk}|^{2}\delta (\omega -\omega _{k})$ are the environmental correlation function and the spectral density, respectively. 
\end{enumerate}

According to these two cases, an arbitrary initial state of the $l$th subsystem $\rho _{\text{tot}}(0)=\rho _{l}(0)\otimes|\{0_{k}\}_{l}\rangle \langle \{0_{k}\}_{l}|$, where $\rho _{l}(0)=\rho_{11}|\uparrow_{l}\rangle \langle \uparrow_{l}|+\rho _{00}|\downarrow_{l}\rangle \langle \downarrow_{l}|+(\rho _{10}|\uparrow_{l}\rangle \langle \downarrow_{l}|+\text{h.c.})$, evolves to
\begin{eqnarray}
\rho _{\text{tot}}(t) &=&\rho _{11}|\Psi _{2}(t)\rangle \langle \Psi_{2}(t)|+\rho _{00}|\downarrow_{l},\{0_{k}\}_{l}\rangle \langle \downarrow_{l},\{0_{k}\}_{l}| \nonumber\\
&&+[\rho _{10}|\Psi _{2}(t)\rangle \langle \downarrow_{l},\{0_{k}\}_{l}|+\text{h.c.}].
\end{eqnarray}
After tracing the environmental states, we have
\begin{eqnarray}
\rho _{l}(t) &=&\rho _{11}|u(t)|^{2}|\uparrow_{l}\rangle \langle \uparrow_{l}|+(1-\rho_{11}|u(t)|^{2}) \nonumber\\
&&  \times|\downarrow_{l}\rangle \langle \downarrow_{l}|+[\rho _{10}u(t)|\uparrow_{l}\rangle \langle \downarrow_{l}|+\text{h.c.}],  \label{smrdcd}
\end{eqnarray}%
where $\rho _{11}+\rho _{00}=1$ has been used. In the basis formed by $|\uparrow_{l}\rangle $ and $|\downarrow_{l}\rangle $, its time derivative is 
\begin{eqnarray}
\dot{\rho}_{l}(t) &=&\Gamma (t)\left(
\begin{array}{cc}
-2\rho _{11}|u(t)|^{2} & -\rho _{10}u(t) \\ 
-\rho _{01}u^{\ast }(t) & 2\rho _{11}|u(t)|^{2}
\end{array}\right)   \nonumber\\
&& -i\Omega (t)\left( 
\begin{array}{cc}
0 & \rho _{10}u(t) \\ 
-\rho _{01}u^{\ast }(t) & 0
\end{array}
\right) ,  \label{densitymatrix2}
\end{eqnarray}
where $\Gamma (t)=-\text{Re}[\frac{\dot{u}(t)}{u(t)}]$ and $\Omega (t)=-\text{Im}[\frac{\dot{u}(t)}{u(t)}]$ denote the decay rate and renormalized frequency of the spin. Using Eq. \eqref{smrdcd} to rewrite $\left( 
\begin{array}{cc}
0 & \rho _{10}u(t) \\ 
-\rho _{01}u^{\ast }(t) & 0
\end{array}
\right) =[\hat{\sigma}_{l+}\hat{\sigma}_{l-},\rho_l (t)]$ and $\left( 
\begin{array}{cc}
-2\rho _{11}|u(t)|^{2} & -\rho _{10}u(t) \\ 
-\rho _{g01}u^{\ast }(t) & 2\rho _{11}|u(t)|^{2}
\end{array}
\right) =2\hat{\sigma}_{l-}\rho_l (t)\hat{\sigma}_{l+}-\hat{\sigma}_{l+}\hat{\sigma}_{l-}\rho_l (t)-\rho_l (t)\hat{\sigma}_{l+}\hat{\sigma}_{l-}\equiv \check{\mathcal{L}}_{l}\rho_l(t)$, we obtain the exact master equation of the $l$th spin as 
\begin{equation}
\dot{\rho}_{l}(t)=-i\Omega (t)[\hat{\sigma}_{l+}\hat{%
\sigma}_{l-},\rho _{l}(t)]+\Gamma (t)\check{\mathcal{L}}_{l}\rho (t). \label{smexactME}
\end{equation}
Because the $N$ subsystems evolve with time independently, we have the master equation of the $N$ spins as
\begin{equation}
\dot{\rho}(t)=\sum_{l=1}^N\{-i\Omega(t)[\hat{\sigma}_{l+}\hat{\sigma}_{l-},\rho(t)]+\Gamma(t)\check{\mathcal L}_l\rho(t)\}.\label{nmeqatsm}
\end{equation}
We find from Eq. \eqref{nmeqatsm} that the non-Markovian decoherence dynamics is essentially determined by the time-dependent $\Omega(t)$ and $\Gamma(t)$. They both are governed by $u(t)$. Equation \eqref{smrdcd} reveals that $u(t)$ is just the time-dependent factor of quantum coherence characterized by the off-diagonal element of the density matrix. Thus, we call it the quantum coherence factor. The solution of Eq. \eqref{nmeqatsm} can be represented by the Kraus operators as $\rho(t)=\check{\Lambda}_t^{\otimes N}\rho(0)$, where $\check{\Lambda}_t\cdot=\sum_{\mu=1}^2\hat{K}_\mu(t)\cdot\hat{K}^\dag_\mu(t)$, $ \hat{K}_{1}=\text{diag}[u(t), 1]$, and $\hat{K}_{2}=[1-|u(t)|^2]^{1/2}\hat{\sigma}_-$. Then, the expectation value for any operator $\hat{O}$ is recast into 
\begin{equation}
\langle \hat{O}\rangle\equiv\text{Tr}[ \hat{O}\check{\Lambda}_t\rho(0)]=\langle(\check{\Lambda}_t^{\dag}\hat{O})\rho(0)\rangle\equiv\langle(\check{\Lambda}_t^{\dag}\hat{O})\rangle_0,\label{exptasm}
\end{equation} which is just the expectation value of $\check{\Lambda}_t^\dag\hat{O}$ in the initial state $\rho(0)$. 

In the special case where the spin-environment coupling is weak and the time scale of the environmental correlation function is much smaller than the one of the spins, we can make the Born-Markov approximation by neglecting the memory effect and extending the upper limit of the integral to $\infty$ \cite{PhysRevLett.102.040403,PhysRevLett.108.130402}. It results in $u_\text{BMA}(t)=e^{-[\kappa+i(\omega_0+\Delta(\omega_0))]t}$ with $\kappa=\pi J(\omega_0)$, $\Delta(\omega_0)=\mathcal{P}\int_{0}^{\infty}\frac{J(\omega)}{\omega_0-\omega}d\omega$, and $\mathcal{P}$ being the Cauchy principal value. Therefore, the quantum coherence irreversibly decays to the environment.

\begin{figure}[t]
\centering
\includegraphics[angle=0,width=.8\columnwidth]{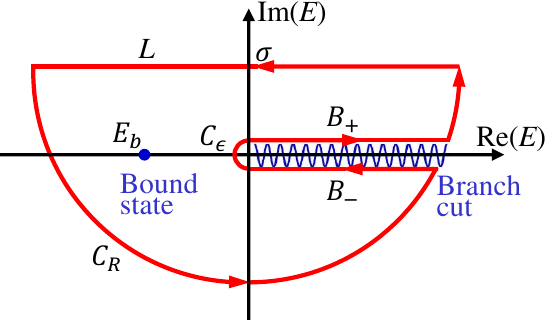}
\caption{ Path of the contour integration in the $\text{Re}(E)$-$\text{Im}(E)$ plane for the calculation of the inverse Laplace transform of $\tilde{u}(E)$.}\label{contin}
\end{figure}

In the general non-Markovian case, the solution of Eq. \eqref{cctd} is formally obtained by Laplace transform, which converts Eq. \eqref{cctd} to $\tilde{u}(z)=[z+i\omega_{0}+\int \frac{J(\omega)}{z+i\omega}d\omega]^{-1}$. $u(t)$ is the inverse Laplace transform of $\tilde{u}(z)$, i.e., $u(t)=\int_{i\sigma+\infty}^{i\sigma-\infty}\tilde{u}(-iE)e^{-iEt}\frac{dE}{2\pi i}$,
where $\sigma$ is larger than any poles of $\tilde{u}(z)$. The poles of $\tilde{u}(z)$ are determined by\begin{equation}
    Y(E)=E,\label{enerSpc}
\end{equation}where $Y(E)\equiv \omega_0-\int_{0}^{\infty}\frac{J(\omega)}{\omega-E}d\omega$. 
It is interesting to see that Eq. \eqref{enerSpc} has an exactly same form as the equation satisfied by the eigenenergies. This signifies that the decoherence dynamics of each spin is essentially determined by the energy-spectrum characteristics of the total spin-environment system. According to the contour integration to evaluate the inverse Laplace transform, see Fig. \ref{contin}, and using the residue theorem, we obtain
\begin{equation}
u(t)=Ze^{-iE_{b}t}+\int_{0}^{\infty}\mathcal{C}(E)e^{-iEt}dE,\label{utana}
\end{equation} 
where $Z=[1+\int_{0}^{\infty}\frac{J(\omega)d\omega}{(E_{b}-\omega)^2}]^{-1}$ originates from the residue contributed by the bound state and $\mathcal{C}(E)=J(E)/\{[E-\omega_0-\Delta(E)]^{2}+[\pi J(E)]^{2}\}$ from the integration paths $B_\pm$ contributed by the continuum energy band. The second term of Eq. \eqref{utana} goes to zero in the long-time limit due to the out-of-phase interference. Thus, we obtain the steady-state solution of Eq. \eqref{utana} as
\begin{equation}
\lim_{t\rightarrow\infty}u(t)=\begin{cases}
0, & Y(0)\geq0\\
Ze^{-iE_{b}t}, & Y(0)<0
\end{cases}.\label{utlong}
\end{equation}

To gain a physical picture of the dynamical consequence of the bound state, we consider again the evolution of $|\Psi(0)\rangle=|\uparrow_l,\{0_k\}\rangle$. Besides expanding as Eq. \eqref{apptmp2}, the evolved state is equivalently expanded in the complete basis formed by the eigenstates of $\hat{H}_l$ as
\begin{equation}
|\Psi(t)\rangle=\alpha_{b}e^{-iE_bt}|\Phi_b\rangle+\sum_{m\in\text{Band}}\beta_m e^{-iE_mt}|\Phi_m\rangle,\label{dvd1}
\end{equation}
where $|\Phi_b\rangle$ is the bound state and $|\Phi_m\rangle$ the eigenstate in the continuum band. The excited-state population of the $l$th spin, i.e., $P(t)=\langle\Psi(t)|\hat{\sigma}_{l+}\hat{\sigma}_{l-}|\Psi(t)\rangle$, calculated from either Eq. \eqref{apptmp2} or Eq. \eqref{dvd1} is
\begin{widetext}
\begin{eqnarray}
P(t)&=&|u(t)|^2=|\alpha_b|^2\langle \Phi_b|\hat{\sigma}_{l+}\hat{\sigma}_{l-}|\Phi_b\rangle+\sum_{m,m'\in \text{Band}}[ \beta^{*}_{m'} \beta_m e^{-i(E_m-E_{m'})t}\nonumber\\
&&\times\langle \Phi_{m'}|\hat{\sigma}_{l+}\hat{\sigma}_{l-}|\Phi_m\rangle+\text{c.c.}]+\sum_{m\in \text{Band}}[\beta^{*}_{m}\alpha_{b}e^{-i(E_b-E_{m})t}\langle \Phi_{m}|\hat{\sigma}_{l+}\hat{\sigma}_{l-}|\Phi_b\rangle+\text{c.c.}].
\end{eqnarray}
\end{widetext}
The second and third terms contain the oscillating frequencies $E_{m}$, which are integrated over the continuum energy band. In the long-time limit, such terms tend to zero due to the destructive interference of different components. Thus, only the contribution of the bound state survives in a long time, i.e.,
\begin{equation}
\lim_{t\rightarrow\infty}P(t)=|\alpha_b|^2\langle \Phi_b|\hat{\sigma}_{l+}\hat{\sigma}_{l-}|\Phi_b\rangle.
\end{equation} It is easy to see from Eq. \eqref{ddeigt} that $\alpha_b=c_b^*$ and $\langle \Phi_b|\hat{\sigma}_{l+}\hat{\sigma}_{l-}|\Phi_b\rangle=|c_b|^2$, where $c_b$ is the excited-state probability amplitude of the bound state $|\Phi_b\rangle$. Then, we have $\lim_{t\rightarrow\infty}P(t)=|c_b|^4$. Furthermore, substituting $d_{kb}={g^*_{lk}c_b\over E_b-\omega_k}$ obtained from the eigenequation for the bound state into the normalization condition $|c_b|^2+\sum_k|d_{kb}|^2=1$, we obtain $|c_b|^2=[ 1+\sum_k{|g_{lk}|^2\over (E_b-\omega_k)^2}]^{-1}$. After taking the continuum limit of the environment frequencies, it becomes
\begin{equation}
   |c_b|^2=[1+\int_0^\infty{J_l(\omega)\over (E_b-\omega)^2}d\omega]^{-1},
\end{equation}which is just $Z$ in Eq. \eqref{utana} obtained from the inverse Laplace transform of $\tilde{u}(z)$. Finally, we obtain a self-consistent result by using two different ways as
\begin{equation}
\lim_{t\rightarrow\infty}P(t)=|c_b|^4=Z^2. 
\end{equation}
Therefore, it is the bound state, which acts as a stationary state and, hence, preserves its component contained in the initial state during the time evolution, that suppresses the decoherence. This is the key physical mechanism of how the bound state suppresses decoherence.

\begin{figure}
\centering
\includegraphics[width=\columnwidth]{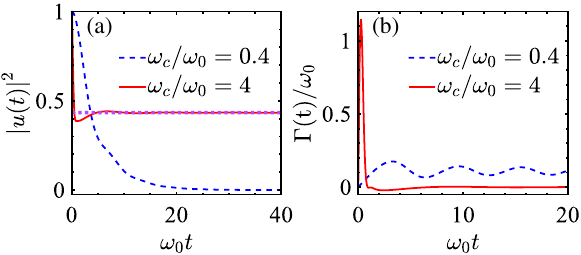}\\
\caption {Evolution of (a) $|u(t)|^2$ and the decay rate $\Gamma(t)=-\text{Re}[\dot{u}(t)/u(t)]$  when $\omega_c=0.4\omega_0$ (blue dashed lines) and $4\omega_0$ (red solid lines). The purple dotted line in (a) is $Z^2$. Other parameters are $\eta=0.5$ and $s=1$.}	\label{fs4}
\end{figure}

Figure \ref{fs4} shows the evolution of $|u(t)|^2$ and the corresponding decay rate $\Gamma(t)$ in Eq. \eqref{nmeqatsm} by numerically calculating Eq. \eqref{cctd}. The result confirms our analytical expectation that $|u(t)|^2$ monotonically decreases to zero and the associated decay rate $\Gamma(t)$ keeps to be positive during the whole time evolution when the bound state is absent; while it tends to $Z^2$ when the bound state is present [see Fig. \ref{fs4}(a)]. The formation of the bound state makes $\Gamma(t)$ transiently being negative values. It reflects the dynamical backflow of quantum coherence from the environment to the system, which is a typical feature of the non-Markovian dynamics. It is remarkable to find from Fig. \ref{fs4}(b) that the formation of the bound state also makes $\Gamma(t)$ tend to zero in the long-time limit.

The result that the poles of the Laplace transform of $u(t)$ is mapped to the eigenenergies is a general property of the total system whose excitation number is conserved. To demonstrate this, we consider the Hamiltonian of an open quantum system interacting with a bosonic environment as $\hat{H}=\hat{H}_S+\hat{H}_E+\hat{H}_I$, where
\begin{eqnarray}\label{totH}
\hat{H}_{S}&=&\omega_0\hat{O}^\dag\hat{O},~~\hat{H}_E=\sum_k\omega_k\hat{a}_k^\dag\hat{a}_k,\\
\hat{H}_I&=&\sum_k(g_k\hat{a}_k\hat{O}^\dag+\text{H.c.}),
\end{eqnarray}
are the Hamiltonians of the system, environment, and their interactions. If the system is a two-level system, then $\hat{O}=\hat{\sigma}_-$. If the system is a harmonic oscillator, then $\hat{O}=\hat{b}$ with $[\hat{b},\hat{b}^\dag]=1$. A common feature of the two situations is that the excitation number is conserved, i.e. $[\hat{H},\hat{\mathcal N}]=0$ with $\hat{\mathcal N}=\hat{O}^\dag\hat{O}+\sum_k\hat{a}_k^\dag\hat{a}_k$. In the above, we have proven that the dynamics of the open two-level system is governed by $u(t)$ satisfying Eq. \eqref{cctd} and the pole of Laplace transform of $u(t)$ can be mapped to the eigenenergies of $\hat{H}$ in the single-excitation subspace. It is interesting to find in Ref. \cite{PhysRevLett.133.050401} that the dynamics of an open harmonic oscillator is also governed by $u(t)$, which has a same form as Eq.  \eqref{cctd} and can be converted into the eigenenergies of $\hat{H}$ in the single-excitation subspace by the Laplace transform. Thus, the non-Markovian dynamics of such kind of open systems is generally determined by the single-excitation eigenenergy spectrum of the total system.

\section{Spin squeezing parameter}\label{sec4}
According to Eq. \eqref{exptasm}, the expectation values of collective spin operators in $\rho(t)$ are converted into the ones of their corresponding inversely evolved operators in $\rho(0)$.  

First, we consider the spin squeezed states from one-axis twisting model
\begin{eqnarray}
   |\Psi_{\text{OAT}}\rangle &=& e^{-i\Theta \hat{J}_{x}^{2}} \,|j,-j\rangle,
\end{eqnarray}with $j=N/2$ and $|j,-j\rangle$ being the common eigenstate of $\hat{J}^2$ and $\hat{J}_z$, as the initial states. Both $|\Psi_\text{OAT}\rangle$ and its evolved state $\rho(t)$ are invariant under the permutation of the constitute spins. Therefore, the expectation values of the collective spin operators satisfy \cite{MA201189}
\begin{eqnarray}
\langle \hat{J}_\alpha\rangle&=&N\langle\hat{\sigma}_{1}^{\alpha}\rangle/2=N\langle\check{\Lambda}^\dag_t\hat{\sigma}_{1}^{\alpha}\rangle_0/2.\label{oattsm1}
\end{eqnarray}
The Kraus representation of $\check{\Lambda}_t$ gives
\begin{eqnarray}
    \check{\Lambda}_t^\dag\hat{\sigma}_{1-}&=& u(t)\hat{\sigma}_{1-},\label{evosm1}\\
    \check{\Lambda}_t^\dag\hat{\sigma}_1^z&=&|u(t)|^2\hat{\sigma}^z_1-1+|u(t)|^2,\label{evosm2}
\end{eqnarray}where $\hat{\sigma}_-=(\hat{\sigma}^x-i\hat{\sigma}^y)/2$. It is straightforward to calculate $\langle\hat{\sigma}_{1-}\rangle_0=0$, which causes $\langle\hat{\sigma}_1^x\rangle_0=\langle\hat{\sigma}_1^y\rangle_0=0$, and $\langle\hat{\sigma}_1^z\rangle_0=-\cos^{2j-1}\Theta$ for $|\Psi_\text{OAT}\rangle$. Then the mean spin is 
\begin{equation}
    \langle{\bf J}\rangle={N\over 2}[|u(t)|^2(1-\cos^{2j-1}\Theta)-1](0,0,1).
\end{equation} The concerned spin component in $\xi$ perpendicular to
$\langle{\bf J}\rangle$ becomes $\hat{J}_{\perp,\beta}=\cos\beta \hat{J}_y+\sin\beta\hat{J}_z$. Its minimal variance by optimizing all $\beta$ reads 
\begin{equation}
    \min_\beta\Delta J_{\perp,\beta}^2=(\langle\hat{J}_x^2+\hat{J}_y^2\rangle-|\langle\hat{J}_-^2\rangle|)/2,\label{mnmsm}
\end{equation} where $\hat{J}_-=\hat{J}_x-i\hat{J}_y$. To calculate the expectation values of the squared spin operators, we again use the  permutation symmetry and obtain 
\begin{eqnarray}
\langle\hat{J}_\alpha^2\rangle&=&N[1+(N-1)\langle \hat{\sigma}_{1}^\alpha\hat{\sigma}_{2}^\alpha\rangle]/4,\nonumber\\
&=&N[1+(N-1)\langle( \check{\Lambda}_t^\dag\hat{\sigma}_{1}^\alpha)( \check{\Lambda}_t^\dag\hat{\sigma}_{2}^\alpha)\rangle_0]/4, \label{tpxy}\\
\langle\hat{J}_-^2\rangle&=&N(N-1)\langle \hat{\sigma}_{1-}\hat{\sigma}_{2-}\rangle\nonumber\\
&=&N(N-1)\langle ( \check{\Lambda}_t^\dag\hat{\sigma}_{1-}) ( \check{\Lambda}_t^\dag\hat{\sigma}_{2-})\rangle_0.~~~~\label{jjmssm}
\end{eqnarray}
The Kraus representations in Eqs. \eqref{evosm1} and \eqref{evosm2} give
\begin{eqnarray}
  && \langle( \check{\Lambda}_t^\dag\hat{\sigma}_{1}^x)( \check{\Lambda}_t^\dag\hat{\sigma}_{2}^x)\rangle_0+\langle( \check{\Lambda}_t^\dag\hat{\sigma}_{1}^y)( \check{\Lambda}_t^\dag\hat{\sigma}_{2}^y)\rangle_0\nonumber\\
   &=&2|u(t)|^2\langle\hat{\sigma}_{1-}\hat{\sigma}_{2+}\rangle_0+\text{c.c.},~~\label{mpsm}\\
   &&\langle ( \check{\Lambda}_t^\dag\hat{\sigma}_{1-})( \check{\Lambda}_t^\dag\hat{\sigma}_{2+})\rangle_0=u(t)^2\langle\hat{\sigma}_{1-}\hat{\sigma}_{2-}\rangle_0.\label{mmsm}
\end{eqnarray}
One can easily calculate $\langle \hat{\sigma}_{1-}\hat{\sigma}_{2+}\rangle_0=A/8$ and $\langle \hat{\sigma}_{1-}\hat{\sigma}_{2-}\rangle_0=-(A-iB)/8$, with $A=1-\cos^{N-2}(2\Theta)$ and $B=4\sin\Theta\cos^{N-2}\Theta$, for $|\Psi_{\text{OAT}}\rangle$. Substituting Eqs. \eqref{mpsm} and \eqref{mmsm} in Eqs. \eqref{tpxy} and \eqref{jjmssm}, we have
\begin{eqnarray}
\min_\beta\Delta J_{\perp,\beta}^2={N\over 4}[1+{(N-1)|u(t)|^2\over 4}(A-\sqrt{A^2+B^2})].\nonumber\\
\end{eqnarray}
Therefore, we finally derive the spin squeezing parameter
\begin{equation}
  \xi_{\rm OAT}^2(t)\simeq1+(N-1)|u(t)|^2(A-\sqrt{A^2+B^2})/4  \label{xismtt}
\end{equation}for the evolved state from the initial state $|\Psi_\text{OAT}\rangle$. 
Accompanying the formation of the bound state, the non-Markovian dynamics recasts Eq. \eqref{xismtt} to the steady-state value
\begin{equation}
 \xi_{\rm OAT}^2(\infty)\simeq1+(N-1)Z^2(A-\sqrt{A^2+B^2})/4 .
\end{equation}
Expanding it in the large-$N$ and small-$\Theta$ limit, i.e., $N\Theta^{2}\ll1\ll N\Theta$, we have \cite{PhysRevA.47.5138}
\begin{equation}
\xi_{\rm OAT}^2(\infty) \simeq 1+Z^{2}(N^{-2}\Theta^{-2}+N^{2}\Theta^{4}/6-1),
\end{equation}
which attains its minimum $\xi_{\rm OAT}^2(\infty)|_{\Theta=\Theta_0}\simeq 1.04Z^2N^{-{2\over 3}}+1-Z^2$ at $\Theta=3^{1/6}N^{2/3}\equiv\Theta_0$. In the large-$N$ limit, we obtain the same $N$-scaling behavior as in the ideal decoherence-free case by making $Z$ as close as possible to one. 

Second, we consider the spin squeezed state from the two-axis twisting model 
\begin{equation}
|\Psi_{\text{TAT}}\rangle=e^{\Theta(\hat{J}_{+}^{2}-\hat{J}_{-}^{2})}\,|j,-j\rangle 
\end{equation}
as the initial state. The nonlinear Hamiltonian counter-twists the collective spin about two orthogonal axes (e.g. $x$ and $y$), analogous to a two-mode parametric amplifier that creates excitations in pairs. It also possesses the permutation symmetry, which enables the calculation of $\langle \hat{J}_\alpha\rangle$ from Eqs. \eqref{oattsm1}, \eqref{evosm1}, and \eqref{evosm2}. The mean spin is $ \langle\hat{\bf J}\rangle=-{N\over2}(0,0,1)$. Thus, the minimal variance in the plane perpendicular to $\langle\hat{\bf J}\rangle$ for the evolved state from the initial two-axis twisted state is the same as the one in Eq. \eqref{mnmsm} for the one-axis twisted state. The use of the bipartite expansion of the expectation values of the squared spin operators in Eqs. \eqref{tpxy} and \eqref{jjmssm} and the Kraus representations in Eqs. \eqref{mpsm} and \eqref{mmsm} results in
\begin{eqnarray}
    \langle\hat{J}_x^2+\hat{J}_y^2\rangle&=&{N[1-|u(t)|^2]\over 2}+|u(t)|^2\langle\hat{J}_x^2+\hat{J}_y^2\rangle_0,~~\\
    \langle\hat{J}_-^2\rangle&=&u(t)^2\langle\hat{J}_-^2\rangle_0,
\end{eqnarray}where the initial conditions $\langle\hat{\sigma}_{1-}\hat{\sigma}_{2+}+\hat{\sigma}_{2-}\hat{\sigma}_{1+}\rangle_0=[2\langle\hat{J}_x^2+\hat{J}_y^2\rangle_0/N-1]/(N-1)$ and $\langle\hat{\sigma}_{1-}\hat{\sigma}_{2-}\rangle_0=\langle\hat{J}_-^2\rangle_0/[N(N-1)]$ from Eqs. \eqref{tpxy} and \eqref{jjmssm} have been used. Then the spin squeezing parameter becomes 
\begin{equation}
\xi_{\text{{TAT}}}^2(t)=1-|u(t)|^{2}+\tfrac{2|u(t)|^{2}}{N}[\langle \hat{J}_{x}^2+\hat{J}_{y}^2 \rangle_{0}-|\langle\hat{J}_{-}^{2}\rangle_{0}|].
\end{equation}
Unlike the one-axis twisted case, the analytical expressions of $\langle \hat{J}_{x}^2+\hat{J}_{y}^2 \rangle_{0}$ and $\langle\hat{J}_{-}^{2}\rangle_{0}$ for $|\Psi_\text{TAT}\rangle$ cannot be obtained. We must resort to numerical calculations.

\section{Husimi Q function}\label{sec5}
The spin coherent state is defined as
\begin{equation}
  |\theta,\varphi\rangle={e^{\zeta \hat{J}_-}|j,j\rangle\over (1+|\zeta|^2)^{j}}=[\cos{\theta\over 2}\left|\uparrow\right\rangle+e^{i\varphi}\sin{\theta\over2}\left|\downarrow\right\rangle]^{\otimes N},  \label{scssm}
\end{equation} where $\zeta=e^{i\varphi}\tan{\theta\over 2}$. The Husimi Q function for the state $\rho(t)$ of the spin system is defined as \begin{eqnarray}
  Q(\theta,\varphi)&=&{2j+1\over 4\pi}\text{Tr}[\rho(t)|\theta,\varphi\rangle\langle\theta,\varphi|]\nonumber\\
  &=&{2j+1\over 4\pi}\langle\Lambda_t^{\dag\otimes N}|\theta,\varphi\rangle\langle\theta,\varphi|\rangle_0,
\end{eqnarray} which maps $\rho(t)$ to a quasiclassical probability distribution in the phase space defined by the spin coherent state $|\theta,\varphi\rangle$. Rewriting Eq. \eqref{scssm} as
\begin{eqnarray} 
  |\theta,\varphi\rangle\langle\theta,\varphi|=\Big[{I\over2}+{\cos\theta\over 2} \hat{\sigma}^z+{\sin\theta\over2}(e^{i\varphi}\hat{\sigma}_-+\text{h.c.})\Big]^{\otimes N},~~~~~
\end{eqnarray} we obtain
\begin{eqnarray}
    Q(\theta,\varphi)&=&{2j+1\over 4\pi}\Big\langle\Big[{I\over 2}+{\cos\theta\over 2}\check{\Lambda}_t^\dag\hat{\sigma}^z\nonumber\\
    &&+{\sin\theta\over 2}(e^{i\varphi}\check{\Lambda}_t^\dag\hat{\sigma}_-+\text{h.c.})\Big]^{\otimes N}\Big\rangle_0.\label{qfctsm}
\end{eqnarray}
Substituting the Kraus representations in Eqs. \eqref{mpsm} and \eqref{mmsm} into Eq. \eqref{qfctsm}, the Husimi Q function of the evolved states from the initial one- and two-axis twisted states can be numerically calculated for given $N$.

\bibliography{references}

\begin{thebibliography}{92}%
\makeatletter
\providecommand \@ifxundefined [1]{%
 \@ifx{#1\undefined}
}%
\providecommand \@ifnum [1]{%
 \ifnum #1\expandafter \@firstoftwo
 \else \expandafter \@secondoftwo
 \fi
}%
\providecommand \@ifx [1]{%
 \ifx #1\expandafter \@firstoftwo
 \else \expandafter \@secondoftwo
 \fi
}%
\providecommand \natexlab [1]{#1}%
\providecommand \enquote  [1]{``#1''}%
\providecommand \bibnamefont  [1]{#1}%
\providecommand \bibfnamefont [1]{#1}%
\providecommand \citenamefont [1]{#1}%
\providecommand \href@noop [0]{\@secondoftwo}%
\providecommand \href [0]{\begingroup \@sanitize@url \@href}%
\providecommand \@href[1]{\@@startlink{#1}\@@href}%
\providecommand \@@href[1]{\endgroup#1\@@endlink}%
\providecommand \@sanitize@url [0]{\catcode `\\12\catcode `\$12\catcode `\&12\catcode `\#12\catcode `\^12\catcode `\_12\catcode `\%12\relax}%
\providecommand \@@startlink[1]{}%
\providecommand \@@endlink[0]{}%
\providecommand \url  [0]{\begingroup\@sanitize@url \@url }%
\providecommand \@url [1]{\endgroup\@href {#1}{\urlprefix }}%
\providecommand \urlprefix  [0]{URL }%
\providecommand \Eprint [0]{\href }%
\providecommand \doibase [0]{https://doi.org/}%
\providecommand \selectlanguage [0]{\@gobble}%
\providecommand \bibinfo  [0]{\@secondoftwo}%
\providecommand \bibfield  [0]{\@secondoftwo}%
\providecommand \translation [1]{[#1]}%
\providecommand \BibitemOpen [0]{}%
\providecommand \bibitemStop [0]{}%
\providecommand \bibitemNoStop [0]{.\EOS\space}%
\providecommand \EOS [0]{\spacefactor3000\relax}%
\providecommand \BibitemShut  [1]{\csname bibitem#1\endcsname}%
\let\auto@bib@innerbib\@empty
\bibitem [{\citenamefont {Degen}\ \emph {et~al.}(2017)\citenamefont {Degen}, \citenamefont {Reinhard},\ and\ \citenamefont {Cappellaro}}]{RevModPhys.89.035002}%
  \BibitemOpen
  \bibfield  {author} {\bibinfo {author} {\bibfnamefont {C.~L.}\ \bibnamefont {Degen}}, \bibinfo {author} {\bibfnamefont {F.}~\bibnamefont {Reinhard}},\ and\ \bibinfo {author} {\bibfnamefont {P.}~\bibnamefont {Cappellaro}},\ }\bibfield  {title} {\bibinfo {title} {Quantum sensing},\ }\href {https://doi.org/10.1103/RevModPhys.89.035002} {\bibfield  {journal} {\bibinfo  {journal} {Rev. Mod. Phys.}\ }\textbf {\bibinfo {volume} {89}},\ \bibinfo {pages} {035002} (\bibinfo {year} {2017})}\BibitemShut {NoStop}%
\bibitem [{\citenamefont {Pezz\`e}\ \emph {et~al.}(2018)\citenamefont {Pezz\`e}, \citenamefont {Smerzi}, \citenamefont {Oberthaler}, \citenamefont {Schmied},\ and\ \citenamefont {Treutlein}}]{RevModPhys.90.035005}%
  \BibitemOpen
  \bibfield  {author} {\bibinfo {author} {\bibfnamefont {L.}~\bibnamefont {Pezz\`e}}, \bibinfo {author} {\bibfnamefont {A.}~\bibnamefont {Smerzi}}, \bibinfo {author} {\bibfnamefont {M.~K.}\ \bibnamefont {Oberthaler}}, \bibinfo {author} {\bibfnamefont {R.}~\bibnamefont {Schmied}},\ and\ \bibinfo {author} {\bibfnamefont {P.}~\bibnamefont {Treutlein}},\ }\bibfield  {title} {\bibinfo {title} {Quantum metrology with nonclassical states of atomic ensembles},\ }\href {https://doi.org/10.1103/RevModPhys.90.035005} {\bibfield  {journal} {\bibinfo  {journal} {Rev. Mod. Phys.}\ }\textbf {\bibinfo {volume} {90}},\ \bibinfo {pages} {035005} (\bibinfo {year} {2018})}\BibitemShut {NoStop}%
\bibitem [{\citenamefont {L\"ucke}\ \emph {et~al.}(2014)\citenamefont {L\"ucke}, \citenamefont {Peise}, \citenamefont {Vitagliano}, \citenamefont {Arlt}, \citenamefont {Santos}, \citenamefont {T\'oth},\ and\ \citenamefont {Klempt}}]{PhysRevLett.112.155304}%
  \BibitemOpen
  \bibfield  {author} {\bibinfo {author} {\bibfnamefont {B.}~\bibnamefont {L\"ucke}}, \bibinfo {author} {\bibfnamefont {J.}~\bibnamefont {Peise}}, \bibinfo {author} {\bibfnamefont {G.}~\bibnamefont {Vitagliano}}, \bibinfo {author} {\bibfnamefont {J.}~\bibnamefont {Arlt}}, \bibinfo {author} {\bibfnamefont {L.}~\bibnamefont {Santos}}, \bibinfo {author} {\bibfnamefont {G.}~\bibnamefont {T\'oth}},\ and\ \bibinfo {author} {\bibfnamefont {C.}~\bibnamefont {Klempt}},\ }\bibfield  {title} {\bibinfo {title} {Detecting multiparticle entanglement of {D}icke states},\ }\href {https://doi.org/10.1103/PhysRevLett.112.155304} {\bibfield  {journal} {\bibinfo  {journal} {Phys. Rev. Lett.}\ }\textbf {\bibinfo {volume} {112}},\ \bibinfo {pages} {155304} (\bibinfo {year} {2014})}\BibitemShut {NoStop}%
\bibitem [{\citenamefont {Wineland}\ \emph {et~al.}(1992)\citenamefont {Wineland}, \citenamefont {Bollinger}, \citenamefont {Itano}, \citenamefont {Moore},\ and\ \citenamefont {Heinzen}}]{PhysRevA.46.R6797}%
  \BibitemOpen
  \bibfield  {author} {\bibinfo {author} {\bibfnamefont {D.~J.}\ \bibnamefont {Wineland}}, \bibinfo {author} {\bibfnamefont {J.~J.}\ \bibnamefont {Bollinger}}, \bibinfo {author} {\bibfnamefont {W.~M.}\ \bibnamefont {Itano}}, \bibinfo {author} {\bibfnamefont {F.~L.}\ \bibnamefont {Moore}},\ and\ \bibinfo {author} {\bibfnamefont {D.~J.}\ \bibnamefont {Heinzen}},\ }\bibfield  {title} {\bibinfo {title} {Spin squeezing and reduced quantum noise in spectroscopy},\ }\href {https://doi.org/10.1103/PhysRevA.46.R6797} {\bibfield  {journal} {\bibinfo  {journal} {Phys. Rev. A}\ }\textbf {\bibinfo {volume} {46}},\ \bibinfo {pages} {R6797} (\bibinfo {year} {1992})}\BibitemShut {NoStop}%
\bibitem [{\citenamefont {Kitagawa}\ and\ \citenamefont {Ueda}(1993)}]{PhysRevA.47.5138}%
  \BibitemOpen
  \bibfield  {author} {\bibinfo {author} {\bibfnamefont {M.}~\bibnamefont {Kitagawa}}\ and\ \bibinfo {author} {\bibfnamefont {M.}~\bibnamefont {Ueda}},\ }\bibfield  {title} {\bibinfo {title} {Squeezed spin states},\ }\href {https://doi.org/10.1103/PhysRevA.47.5138} {\bibfield  {journal} {\bibinfo  {journal} {Phys. Rev. A}\ }\textbf {\bibinfo {volume} {47}},\ \bibinfo {pages} {5138} (\bibinfo {year} {1993})}\BibitemShut {NoStop}%
\bibitem [{\citenamefont {Wineland}\ \emph {et~al.}(1994)\citenamefont {Wineland}, \citenamefont {Bollinger}, \citenamefont {Itano},\ and\ \citenamefont {Heinzen}}]{PhysRevA.50.67}%
  \BibitemOpen
  \bibfield  {author} {\bibinfo {author} {\bibfnamefont {D.~J.}\ \bibnamefont {Wineland}}, \bibinfo {author} {\bibfnamefont {J.~J.}\ \bibnamefont {Bollinger}}, \bibinfo {author} {\bibfnamefont {W.~M.}\ \bibnamefont {Itano}},\ and\ \bibinfo {author} {\bibfnamefont {D.~J.}\ \bibnamefont {Heinzen}},\ }\bibfield  {title} {\bibinfo {title} {Squeezed atomic states and projection noise in spectroscopy},\ }\href {https://doi.org/10.1103/PhysRevA.50.67} {\bibfield  {journal} {\bibinfo  {journal} {Phys. Rev. A}\ }\textbf {\bibinfo {volume} {50}},\ \bibinfo {pages} {67} (\bibinfo {year} {1994})}\BibitemShut {NoStop}%
\bibitem [{\citenamefont {Ma}\ \emph {et~al.}(2011)\citenamefont {Ma}, \citenamefont {Wang}, \citenamefont {Sun},\ and\ \citenamefont {Nori}}]{MA201189}%
  \BibitemOpen
  \bibfield  {author} {\bibinfo {author} {\bibfnamefont {J.}~\bibnamefont {Ma}}, \bibinfo {author} {\bibfnamefont {X.}~\bibnamefont {Wang}}, \bibinfo {author} {\bibfnamefont {C.}~\bibnamefont {Sun}},\ and\ \bibinfo {author} {\bibfnamefont {F.}~\bibnamefont {Nori}},\ }\bibfield  {title} {\bibinfo {title} {Quantum spin squeezing},\ }\href {https://doi.org/https://doi.org/10.1016/j.physrep.2011.08.003} {\bibfield  {journal} {\bibinfo  {journal} {Physics Reports}\ }\textbf {\bibinfo {volume} {509}},\ \bibinfo {pages} {89} (\bibinfo {year} {2011})}\BibitemShut {NoStop}%
\bibitem [{\citenamefont {Rosi}\ \emph {et~al.}(2014)\citenamefont {Rosi}, \citenamefont {Sorrentino}, \citenamefont {Cacciapuoti}, \citenamefont {Prevedelli},\ and\ \citenamefont {Tino}}]{Rosi2014}%
  \BibitemOpen
  \bibfield  {author} {\bibinfo {author} {\bibfnamefont {G.}~\bibnamefont {Rosi}}, \bibinfo {author} {\bibfnamefont {F.}~\bibnamefont {Sorrentino}}, \bibinfo {author} {\bibfnamefont {L.}~\bibnamefont {Cacciapuoti}}, \bibinfo {author} {\bibfnamefont {M.}~\bibnamefont {Prevedelli}},\ and\ \bibinfo {author} {\bibfnamefont {G.~M.}\ \bibnamefont {Tino}},\ }\bibfield  {title} {\bibinfo {title} {Precision measurement of the {n}ewtonian gravitational constant using cold atoms},\ }\href {https://doi.org/10.1038/nature13433} {\bibfield  {journal} {\bibinfo  {journal} {Nature}\ }\textbf {\bibinfo {volume} {510}},\ \bibinfo {pages} {518} (\bibinfo {year} {2014})}\BibitemShut {NoStop}%
\bibitem [{\citenamefont {Garrido~Alzar}(2019)}]{10.1116/1.5120348}%
  \BibitemOpen
  \bibfield  {author} {\bibinfo {author} {\bibfnamefont {C.~L.}\ \bibnamefont {Garrido~Alzar}},\ }\bibfield  {title} {\bibinfo {title} {Compact chip-scale guided cold atom gyrometers for inertial navigation: Enabling technologies and design study},\ }\href {https://doi.org/10.1116/1.5120348} {\bibfield  {journal} {\bibinfo  {journal} {AVS Quantum Science}\ }\textbf {\bibinfo {volume} {1}},\ \bibinfo {pages} {014702} (\bibinfo {year} {2019})}\BibitemShut {NoStop}%
\bibitem [{\citenamefont {Salvi}\ \emph {et~al.}(2018)\citenamefont {Salvi}, \citenamefont {Poli}, \citenamefont {Vuleti\ifmmode~\acute{c}\else \'{c}\fi{}},\ and\ \citenamefont {Tino}}]{PhysRevLett.120.033601}%
  \BibitemOpen
  \bibfield  {author} {\bibinfo {author} {\bibfnamefont {L.}~\bibnamefont {Salvi}}, \bibinfo {author} {\bibfnamefont {N.}~\bibnamefont {Poli}}, \bibinfo {author} {\bibfnamefont {V.}~\bibnamefont {Vuleti\ifmmode~\acute{c}\else \'{c}\fi{}}},\ and\ \bibinfo {author} {\bibfnamefont {G.~M.}\ \bibnamefont {Tino}},\ }\bibfield  {title} {\bibinfo {title} {Squeezing on momentum states for atom interferometry},\ }\href {https://doi.org/10.1103/PhysRevLett.120.033601} {\bibfield  {journal} {\bibinfo  {journal} {Phys. Rev. Lett.}\ }\textbf {\bibinfo {volume} {120}},\ \bibinfo {pages} {033601} (\bibinfo {year} {2018})}\BibitemShut {NoStop}%
\bibitem [{\citenamefont {Szigeti}\ \emph {et~al.}(2020)\citenamefont {Szigeti}, \citenamefont {Nolan}, \citenamefont {Close},\ and\ \citenamefont {Haine}}]{PhysRevLett.125.100402}%
  \BibitemOpen
  \bibfield  {author} {\bibinfo {author} {\bibfnamefont {S.~S.}\ \bibnamefont {Szigeti}}, \bibinfo {author} {\bibfnamefont {S.~P.}\ \bibnamefont {Nolan}}, \bibinfo {author} {\bibfnamefont {J.~D.}\ \bibnamefont {Close}},\ and\ \bibinfo {author} {\bibfnamefont {S.~A.}\ \bibnamefont {Haine}},\ }\bibfield  {title} {\bibinfo {title} {High-precision quantum-enhanced gravimetry with a {B}ose-{E}instein condensate},\ }\href {https://doi.org/10.1103/PhysRevLett.125.100402} {\bibfield  {journal} {\bibinfo  {journal} {Phys. Rev. Lett.}\ }\textbf {\bibinfo {volume} {125}},\ \bibinfo {pages} {100402} (\bibinfo {year} {2020})}\BibitemShut {NoStop}%
\bibitem [{\citenamefont {Gil}\ \emph {et~al.}(2014)\citenamefont {Gil}, \citenamefont {Mukherjee}, \citenamefont {Bridge}, \citenamefont {Jones},\ and\ \citenamefont {Pohl}}]{PhysRevLett.112.103601}%
  \BibitemOpen
  \bibfield  {author} {\bibinfo {author} {\bibfnamefont {L.~I.~R.}\ \bibnamefont {Gil}}, \bibinfo {author} {\bibfnamefont {R.}~\bibnamefont {Mukherjee}}, \bibinfo {author} {\bibfnamefont {E.~M.}\ \bibnamefont {Bridge}}, \bibinfo {author} {\bibfnamefont {M.~P.~A.}\ \bibnamefont {Jones}},\ and\ \bibinfo {author} {\bibfnamefont {T.}~\bibnamefont {Pohl}},\ }\bibfield  {title} {\bibinfo {title} {Spin squeezing in a {R}ydberg lattice clock},\ }\href {https://doi.org/10.1103/PhysRevLett.112.103601} {\bibfield  {journal} {\bibinfo  {journal} {Phys. Rev. Lett.}\ }\textbf {\bibinfo {volume} {112}},\ \bibinfo {pages} {103601} (\bibinfo {year} {2014})}\BibitemShut {NoStop}%
\bibitem [{\citenamefont {Pezz\`e}\ and\ \citenamefont {Smerzi}(2020)}]{PhysRevLett.125.210503}%
  \BibitemOpen
  \bibfield  {author} {\bibinfo {author} {\bibfnamefont {L.}~\bibnamefont {Pezz\`e}}\ and\ \bibinfo {author} {\bibfnamefont {A.}~\bibnamefont {Smerzi}},\ }\bibfield  {title} {\bibinfo {title} {Heisenberg-limited noisy atomic clock using a hybrid coherent and squeezed state protocol},\ }\href {https://doi.org/10.1103/PhysRevLett.125.210503} {\bibfield  {journal} {\bibinfo  {journal} {Phys. Rev. Lett.}\ }\textbf {\bibinfo {volume} {125}},\ \bibinfo {pages} {210503} (\bibinfo {year} {2020})}\BibitemShut {NoStop}%
\bibitem [{\citenamefont {Schulte}\ \emph {et~al.}(2020{\natexlab{a}})\citenamefont {Schulte}, \citenamefont {Lisdat}, \citenamefont {Schmidt}, \citenamefont {Sterr},\ and\ \citenamefont {Hammerer}}]{Schulte2020}%
  \BibitemOpen
  \bibfield  {author} {\bibinfo {author} {\bibfnamefont {M.}~\bibnamefont {Schulte}}, \bibinfo {author} {\bibfnamefont {C.}~\bibnamefont {Lisdat}}, \bibinfo {author} {\bibfnamefont {P.~O.}\ \bibnamefont {Schmidt}}, \bibinfo {author} {\bibfnamefont {U.}~\bibnamefont {Sterr}},\ and\ \bibinfo {author} {\bibfnamefont {K.}~\bibnamefont {Hammerer}},\ }\bibfield  {title} {\bibinfo {title} {Prospects and challenges for squeezing-enhanced optical atomic clocks},\ }\href {https://doi.org/10.1038/s41467-020-19403-7} {\bibfield  {journal} {\bibinfo  {journal} {Nat. Commun.}\ }\textbf {\bibinfo {volume} {11}},\ \bibinfo {pages} {5955} (\bibinfo {year} {2020}{\natexlab{a}})}\BibitemShut {NoStop}%
\bibitem [{\citenamefont {Malia}\ \emph {et~al.}(2022)\citenamefont {Malia}, \citenamefont {Wu}, \citenamefont {Martínez-Rincón},\ and\ \citenamefont {Kasevich}}]{Malia2022}%
  \BibitemOpen
  \bibfield  {author} {\bibinfo {author} {\bibfnamefont {B.~K.}\ \bibnamefont {Malia}}, \bibinfo {author} {\bibfnamefont {Y.}~\bibnamefont {Wu}}, \bibinfo {author} {\bibfnamefont {J.}~\bibnamefont {Martínez-Rincón}},\ and\ \bibinfo {author} {\bibfnamefont {M.~A.}\ \bibnamefont {Kasevich}},\ }\bibfield  {title} {\bibinfo {title} {Distributed quantum sensing with mode-entangled spin-squeezed atomic states},\ }\href {https://doi.org/10.1038/s41586-022-05363-z} {\bibfield  {journal} {\bibinfo  {journal} {Nature}\ }\textbf {\bibinfo {volume} {612}},\ \bibinfo {pages} {661} (\bibinfo {year} {2022})}\BibitemShut {NoStop}%
\bibitem [{\citenamefont {Eckner}\ \emph {et~al.}(2023)\citenamefont {Eckner}, \citenamefont {Oppong}, \citenamefont {Cao}, \citenamefont {Young}, \citenamefont {Milner}, \citenamefont {Robinson}, \citenamefont {Ye},\ and\ \citenamefont {Kaufman}}]{Eckner2023}%
  \BibitemOpen
  \bibfield  {author} {\bibinfo {author} {\bibfnamefont {W.~J.}\ \bibnamefont {Eckner}}, \bibinfo {author} {\bibfnamefont {N.~D.}\ \bibnamefont {Oppong}}, \bibinfo {author} {\bibfnamefont {A.}~\bibnamefont {Cao}}, \bibinfo {author} {\bibfnamefont {A.~W.}\ \bibnamefont {Young}}, \bibinfo {author} {\bibfnamefont {W.~R.}\ \bibnamefont {Milner}}, \bibinfo {author} {\bibfnamefont {J.~M.}\ \bibnamefont {Robinson}}, \bibinfo {author} {\bibfnamefont {J.}~\bibnamefont {Ye}},\ and\ \bibinfo {author} {\bibfnamefont {A.~M.}\ \bibnamefont {Kaufman}},\ }\bibfield  {title} {\bibinfo {title} {Realizing spin squeezing with {R}ydberg interactions in an optical clock},\ }\href {https://doi.org/10.1038/s41586-023-06360-6} {\bibfield  {journal} {\bibinfo  {journal} {Nature}\ }\textbf {\bibinfo {volume} {621}},\ \bibinfo {pages} {734} (\bibinfo {year} {2023})}\BibitemShut {NoStop}%
\bibitem [{\citenamefont {Robinson}\ \emph {et~al.}(2024)\citenamefont {Robinson}, \citenamefont {Miklos}, \citenamefont {Tso}, \citenamefont {Kennedy}, \citenamefont {Bothwell}, \citenamefont {Kedar}, \citenamefont {Thompson},\ and\ \citenamefont {Ye}}]{Robinson2024}%
  \BibitemOpen
  \bibfield  {author} {\bibinfo {author} {\bibfnamefont {J.~M.}\ \bibnamefont {Robinson}}, \bibinfo {author} {\bibfnamefont {M.}~\bibnamefont {Miklos}}, \bibinfo {author} {\bibfnamefont {Y.~M.}\ \bibnamefont {Tso}}, \bibinfo {author} {\bibfnamefont {C.~J.}\ \bibnamefont {Kennedy}}, \bibinfo {author} {\bibfnamefont {T.}~\bibnamefont {Bothwell}}, \bibinfo {author} {\bibfnamefont {D.}~\bibnamefont {Kedar}}, \bibinfo {author} {\bibfnamefont {J.~K.}\ \bibnamefont {Thompson}},\ and\ \bibinfo {author} {\bibfnamefont {J.}~\bibnamefont {Ye}},\ }\bibfield  {title} {\bibinfo {title} {Direct comparison of two spin-squeezed optical clock ensembles at the \(10^{-17}\) level},\ }\href {https://doi.org/10.1038/s41567-023-02310-1} {\bibfield  {journal} {\bibinfo  {journal} {Nature Physics}\ }\textbf {\bibinfo {volume} {20}},\ \bibinfo {pages} {208} (\bibinfo {year} {2024})}\BibitemShut {NoStop}%
\bibitem [{\citenamefont {Muessel}\ \emph {et~al.}(2014)\citenamefont {Muessel}, \citenamefont {Strobel}, \citenamefont {Linnemann}, \citenamefont {Hume},\ and\ \citenamefont {Oberthaler}}]{PhysRevLett.113.103004}%
  \BibitemOpen
  \bibfield  {author} {\bibinfo {author} {\bibfnamefont {W.}~\bibnamefont {Muessel}}, \bibinfo {author} {\bibfnamefont {H.}~\bibnamefont {Strobel}}, \bibinfo {author} {\bibfnamefont {D.}~\bibnamefont {Linnemann}}, \bibinfo {author} {\bibfnamefont {D.~B.}\ \bibnamefont {Hume}},\ and\ \bibinfo {author} {\bibfnamefont {M.~K.}\ \bibnamefont {Oberthaler}},\ }\bibfield  {title} {\bibinfo {title} {Scalable spin squeezing for quantum-enhanced magnetometry with {B}ose-{E}instein condensates},\ }\href {https://doi.org/10.1103/PhysRevLett.113.103004} {\bibfield  {journal} {\bibinfo  {journal} {Phys. Rev. Lett.}\ }\textbf {\bibinfo {volume} {113}},\ \bibinfo {pages} {103004} (\bibinfo {year} {2014})}\BibitemShut {NoStop}%
\bibitem [{\citenamefont {Brask}\ \emph {et~al.}(2015)\citenamefont {Brask}, \citenamefont {Chaves},\ and\ \citenamefont {Ko\l{}ody\ifmmode~\acute{n}\else \'{n}\fi{}ski}}]{PhysRevX.5.031010}%
  \BibitemOpen
  \bibfield  {author} {\bibinfo {author} {\bibfnamefont {J.~B.}\ \bibnamefont {Brask}}, \bibinfo {author} {\bibfnamefont {R.}~\bibnamefont {Chaves}},\ and\ \bibinfo {author} {\bibfnamefont {J.}~\bibnamefont {Ko\l{}ody\ifmmode~\acute{n}\else \'{n}\fi{}ski}},\ }\bibfield  {title} {\bibinfo {title} {Improved quantum magnetometry beyond the standard quantum limit},\ }\href {https://doi.org/10.1103/PhysRevX.5.031010} {\bibfield  {journal} {\bibinfo  {journal} {Phys. Rev. X}\ }\textbf {\bibinfo {volume} {5}},\ \bibinfo {pages} {031010} (\bibinfo {year} {2015})}\BibitemShut {NoStop}%
\bibitem [{\citenamefont {Bao}\ \emph {et~al.}(2020)\citenamefont {Bao}, \citenamefont {Duan}, \citenamefont {Jin}, \citenamefont {Lu}, \citenamefont {Li}, \citenamefont {Qu}, \citenamefont {Wang}, \citenamefont {Novikova}, \citenamefont {Mikhailov}, \citenamefont {Zhao}, \citenamefont {M$\o$lmer}, \citenamefont {Shen},\ and\ \citenamefont {Xiao}}]{Bao2020}%
  \BibitemOpen
  \bibfield  {author} {\bibinfo {author} {\bibfnamefont {H.}~\bibnamefont {Bao}}, \bibinfo {author} {\bibfnamefont {J.}~\bibnamefont {Duan}}, \bibinfo {author} {\bibfnamefont {S.}~\bibnamefont {Jin}}, \bibinfo {author} {\bibfnamefont {X.}~\bibnamefont {Lu}}, \bibinfo {author} {\bibfnamefont {P.}~\bibnamefont {Li}}, \bibinfo {author} {\bibfnamefont {W.}~\bibnamefont {Qu}}, \bibinfo {author} {\bibfnamefont {M.}~\bibnamefont {Wang}}, \bibinfo {author} {\bibfnamefont {I.}~\bibnamefont {Novikova}}, \bibinfo {author} {\bibfnamefont {E.~E.}\ \bibnamefont {Mikhailov}}, \bibinfo {author} {\bibfnamefont {K.-F.}\ \bibnamefont {Zhao}}, \bibinfo {author} {\bibfnamefont {K.}~\bibnamefont {M$\o$lmer}}, \bibinfo {author} {\bibfnamefont {H.}~\bibnamefont {Shen}},\ and\ \bibinfo {author} {\bibfnamefont {Y.}~\bibnamefont {Xiao}},\ }\bibfield  {title} {\bibinfo {title} {Spin squeezing of $10^{11}$ atoms by prediction and retrodiction measurements},\ }\href {https://doi.org/10.1038/s41586-020-2243-7} {\bibfield  {journal}
  {\bibinfo  {journal} {Nature}\ }\textbf {\bibinfo {volume} {581}},\ \bibinfo {pages} {159} (\bibinfo {year} {2020})}\BibitemShut {NoStop}%
\bibitem [{\citenamefont {Troullinou}\ \emph {et~al.}(2021)\citenamefont {Troullinou}, \citenamefont {Jim\'enez-Mart\'{\i}nez}, \citenamefont {Kong}, \citenamefont {Lucivero},\ and\ \citenamefont {Mitchell}}]{PhysRevLett.127.193601}%
  \BibitemOpen
  \bibfield  {author} {\bibinfo {author} {\bibfnamefont {C.}~\bibnamefont {Troullinou}}, \bibinfo {author} {\bibfnamefont {R.}~\bibnamefont {Jim\'enez-Mart\'{\i}nez}}, \bibinfo {author} {\bibfnamefont {J.}~\bibnamefont {Kong}}, \bibinfo {author} {\bibfnamefont {V.~G.}\ \bibnamefont {Lucivero}},\ and\ \bibinfo {author} {\bibfnamefont {M.~W.}\ \bibnamefont {Mitchell}},\ }\bibfield  {title} {\bibinfo {title} {Squeezed-light enhancement and backaction evasion in a high sensitivity optically pumped magnetometer},\ }\href {https://doi.org/10.1103/PhysRevLett.127.193601} {\bibfield  {journal} {\bibinfo  {journal} {Phys. Rev. Lett.}\ }\textbf {\bibinfo {volume} {127}},\ \bibinfo {pages} {193601} (\bibinfo {year} {2021})}\BibitemShut {NoStop}%
\bibitem [{\citenamefont {Jin}\ \emph {et~al.}(2021)\citenamefont {Jin}, \citenamefont {Bao}, \citenamefont {Duan}, \citenamefont {Lu}, \citenamefont {Wang}, \citenamefont {Zhao}, \citenamefont {Shen},\ and\ \citenamefont {Xiao}}]{Jin:21}%
  \BibitemOpen
  \bibfield  {author} {\bibinfo {author} {\bibfnamefont {S.}~\bibnamefont {Jin}}, \bibinfo {author} {\bibfnamefont {H.}~\bibnamefont {Bao}}, \bibinfo {author} {\bibfnamefont {J.}~\bibnamefont {Duan}}, \bibinfo {author} {\bibfnamefont {X.}~\bibnamefont {Lu}}, \bibinfo {author} {\bibfnamefont {M.}~\bibnamefont {Wang}}, \bibinfo {author} {\bibfnamefont {K.-F.}\ \bibnamefont {Zhao}}, \bibinfo {author} {\bibfnamefont {H.}~\bibnamefont {Shen}},\ and\ \bibinfo {author} {\bibfnamefont {Y.}~\bibnamefont {Xiao}},\ }\bibfield  {title} {\bibinfo {title} {Adiabaticity in state preparation for spin squeezing of large atom ensembles},\ }\href {https://doi.org/10.1364/PRJ.413288} {\bibfield  {journal} {\bibinfo  {journal} {Photon. Res.}\ }\textbf {\bibinfo {volume} {9}},\ \bibinfo {pages} {2296} (\bibinfo {year} {2021})}\BibitemShut {NoStop}%
\bibitem [{\citenamefont {Kong}\ \emph {et~al.}(2020)\citenamefont {Kong}, \citenamefont {Jiménez-Martínez}, \citenamefont {Troullinou}, \citenamefont {Lucivero}, \citenamefont {Tóth},\ and\ \citenamefont {Mitchell}}]{Kong2020}%
  \BibitemOpen
  \bibfield  {author} {\bibinfo {author} {\bibfnamefont {J.}~\bibnamefont {Kong}}, \bibinfo {author} {\bibfnamefont {R.}~\bibnamefont {Jiménez-Martínez}}, \bibinfo {author} {\bibfnamefont {C.}~\bibnamefont {Troullinou}}, \bibinfo {author} {\bibfnamefont {V.~G.}\ \bibnamefont {Lucivero}}, \bibinfo {author} {\bibfnamefont {G.}~\bibnamefont {Tóth}},\ and\ \bibinfo {author} {\bibfnamefont {M.~W.}\ \bibnamefont {Mitchell}},\ }\bibfield  {title} {\bibinfo {title} {Measurement-induced, spatially-extended entanglement in a hot, strongly-interacting atomic system},\ }\href {https://doi.org/10.1038/s41467-020-15899-1} {\bibfield  {journal} {\bibinfo  {journal} {Nature Communications}\ }\textbf {\bibinfo {volume} {11}},\ \bibinfo {pages} {2415} (\bibinfo {year} {2020})}\BibitemShut {NoStop}%
\bibitem [{\citenamefont {Wang}\ \emph {et~al.}(2010)\citenamefont {Wang}, \citenamefont {Miranowicz}, \citenamefont {Liu}, \citenamefont {Sun},\ and\ \citenamefont {Nori}}]{PhysRevA.81.022106}%
  \BibitemOpen
  \bibfield  {author} {\bibinfo {author} {\bibfnamefont {X.}~\bibnamefont {Wang}}, \bibinfo {author} {\bibfnamefont {A.}~\bibnamefont {Miranowicz}}, \bibinfo {author} {\bibfnamefont {Y.-x.}\ \bibnamefont {Liu}}, \bibinfo {author} {\bibfnamefont {C.~P.}\ \bibnamefont {Sun}},\ and\ \bibinfo {author} {\bibfnamefont {F.}~\bibnamefont {Nori}},\ }\bibfield  {title} {\bibinfo {title} {Sudden vanishing of spin squeezing under decoherence},\ }\href {https://doi.org/10.1103/PhysRevA.81.022106} {\bibfield  {journal} {\bibinfo  {journal} {Phys. Rev. A}\ }\textbf {\bibinfo {volume} {81}},\ \bibinfo {pages} {022106} (\bibinfo {year} {2010})}\BibitemShut {NoStop}%
\bibitem [{\citenamefont {Xue}(2012)}]{PhysRevA.86.023812}%
  \BibitemOpen
  \bibfield  {author} {\bibinfo {author} {\bibfnamefont {P.}~\bibnamefont {Xue}},\ }\bibfield  {title} {\bibinfo {title} {Spin-squeezing property of weighted graph states},\ }\href {https://doi.org/10.1103/PhysRevA.86.023812} {\bibfield  {journal} {\bibinfo  {journal} {Phys. Rev. A}\ }\textbf {\bibinfo {volume} {86}},\ \bibinfo {pages} {023812} (\bibinfo {year} {2012})}\BibitemShut {NoStop}%
\bibitem [{\citenamefont {Koppenh\"ofer}\ and\ \citenamefont {Clerk}(2023)}]{PhysRevResearch.5.043279}%
  \BibitemOpen
  \bibfield  {author} {\bibinfo {author} {\bibfnamefont {M.}~\bibnamefont {Koppenh\"ofer}}\ and\ \bibinfo {author} {\bibfnamefont {A.~A.}\ \bibnamefont {Clerk}},\ }\bibfield  {title} {\bibinfo {title} {Revisiting the impact of dissipation on time-reversed one-axis-twist quantum-sensing protocols},\ }\href {https://doi.org/10.1103/PhysRevResearch.5.043279} {\bibfield  {journal} {\bibinfo  {journal} {Phys. Rev. Res.}\ }\textbf {\bibinfo {volume} {5}},\ \bibinfo {pages} {043279} (\bibinfo {year} {2023})}\BibitemShut {NoStop}%
\bibitem [{\citenamefont {Jiao}\ \emph {et~al.}(2025)\citenamefont {Jiao}, \citenamefont {Wu}, \citenamefont {Bai},\ and\ \citenamefont {An}}]{https://doi.org/10.1002/qute.202300218}%
  \BibitemOpen
  \bibfield  {author} {\bibinfo {author} {\bibfnamefont {L.}~\bibnamefont {Jiao}}, \bibinfo {author} {\bibfnamefont {W.}~\bibnamefont {Wu}}, \bibinfo {author} {\bibfnamefont {S.-Y.}\ \bibnamefont {Bai}},\ and\ \bibinfo {author} {\bibfnamefont {J.-H.}\ \bibnamefont {An}},\ }\bibfield  {title} {\bibinfo {title} {Quantum metrology in the noisy intermediate-scale quantum era},\ }\href {https://doi.org/https://doi.org/10.1002/qute.202300218} {\bibfield  {journal} {\bibinfo  {journal} {Advanced Quantum Technologies}\ }\textbf {\bibinfo {volume} {8}},\ \bibinfo {pages} {2300218} (\bibinfo {year} {2025})}\BibitemShut {NoStop}%
\bibitem [{\citenamefont {Huang}\ \emph {et~al.}(2023{\natexlab{a}})\citenamefont {Huang}, \citenamefont {de~la Paz}, \citenamefont {Mazzoni}, \citenamefont {Ott}, \citenamefont {Rosenbusch}, \citenamefont {Sinatra}, \citenamefont {Garrido~Alzar},\ and\ \citenamefont {Reichel}}]{PRXQuantum.4.020322}%
  \BibitemOpen
  \bibfield  {author} {\bibinfo {author} {\bibfnamefont {M.-Z.}\ \bibnamefont {Huang}}, \bibinfo {author} {\bibfnamefont {J.~A.}\ \bibnamefont {de~la Paz}}, \bibinfo {author} {\bibfnamefont {T.}~\bibnamefont {Mazzoni}}, \bibinfo {author} {\bibfnamefont {K.}~\bibnamefont {Ott}}, \bibinfo {author} {\bibfnamefont {P.}~\bibnamefont {Rosenbusch}}, \bibinfo {author} {\bibfnamefont {A.}~\bibnamefont {Sinatra}}, \bibinfo {author} {\bibfnamefont {C.~L.}\ \bibnamefont {Garrido~Alzar}},\ and\ \bibinfo {author} {\bibfnamefont {J.}~\bibnamefont {Reichel}},\ }\bibfield  {title} {\bibinfo {title} {Observing spin-squeezed states under spin-exchange collisions for a second},\ }\href {https://doi.org/10.1103/PRXQuantum.4.020322} {\bibfield  {journal} {\bibinfo  {journal} {PRX Quantum}\ }\textbf {\bibinfo {volume} {4}},\ \bibinfo {pages} {020322} (\bibinfo {year} {2023}{\natexlab{a}})}\BibitemShut {NoStop}%
\bibitem [{\citenamefont {Baamara}\ \emph {et~al.}(2021)\citenamefont {Baamara}, \citenamefont {Sinatra},\ and\ \citenamefont {Gessner}}]{PhysRevLett.127.160501}%
  \BibitemOpen
  \bibfield  {author} {\bibinfo {author} {\bibfnamefont {Y.}~\bibnamefont {Baamara}}, \bibinfo {author} {\bibfnamefont {A.}~\bibnamefont {Sinatra}},\ and\ \bibinfo {author} {\bibfnamefont {M.}~\bibnamefont {Gessner}},\ }\bibfield  {title} {\bibinfo {title} {Scaling laws for the sensitivity enhancement of non-{G}aussian spin states},\ }\href {https://doi.org/10.1103/PhysRevLett.127.160501} {\bibfield  {journal} {\bibinfo  {journal} {Phys. Rev. Lett.}\ }\textbf {\bibinfo {volume} {127}},\ \bibinfo {pages} {160501} (\bibinfo {year} {2021})}\BibitemShut {NoStop}%
\bibitem [{\citenamefont {Bornet}\ \emph {et~al.}(2023)\citenamefont {Bornet}, \citenamefont {Emperauger}, \citenamefont {Chen}, \citenamefont {Ye}, \citenamefont {Block}, \citenamefont {Bintz}, \citenamefont {Boyd}, \citenamefont {Barredo}, \citenamefont {Comparin}, \citenamefont {Mezzacapo}, \citenamefont {Roscilde}, \citenamefont {Lahaye}, \citenamefont {Yao},\ and\ \citenamefont {Browaeys}}]{Bornet2023}%
  \BibitemOpen
  \bibfield  {author} {\bibinfo {author} {\bibfnamefont {G.}~\bibnamefont {Bornet}}, \bibinfo {author} {\bibfnamefont {G.}~\bibnamefont {Emperauger}}, \bibinfo {author} {\bibfnamefont {C.}~\bibnamefont {Chen}}, \bibinfo {author} {\bibfnamefont {B.}~\bibnamefont {Ye}}, \bibinfo {author} {\bibfnamefont {M.}~\bibnamefont {Block}}, \bibinfo {author} {\bibfnamefont {M.}~\bibnamefont {Bintz}}, \bibinfo {author} {\bibfnamefont {J.~A.}\ \bibnamefont {Boyd}}, \bibinfo {author} {\bibfnamefont {D.}~\bibnamefont {Barredo}}, \bibinfo {author} {\bibfnamefont {T.}~\bibnamefont {Comparin}}, \bibinfo {author} {\bibfnamefont {F.}~\bibnamefont {Mezzacapo}}, \bibinfo {author} {\bibfnamefont {T.}~\bibnamefont {Roscilde}}, \bibinfo {author} {\bibfnamefont {T.}~\bibnamefont {Lahaye}}, \bibinfo {author} {\bibfnamefont {N.~Y.}\ \bibnamefont {Yao}},\ and\ \bibinfo {author} {\bibfnamefont {A.}~\bibnamefont {Browaeys}},\ }\bibfield  {title} {\bibinfo {title} {Scalable spin squeezing in a dipolar {R}ydberg atom array},\ }\href
  {https://doi.org/10.1038/s41586-023-06414-9} {\bibfield  {journal} {\bibinfo  {journal} {Nature}\ }\textbf {\bibinfo {volume} {621}},\ \bibinfo {pages} {728} (\bibinfo {year} {2023})}\BibitemShut {NoStop}%
\bibitem [{\citenamefont {Roscilde}\ \emph {et~al.}(2024)\citenamefont {Roscilde}, \citenamefont {Caleca}, \citenamefont {Angelone},\ and\ \citenamefont {Mezzacapo}}]{PhysRevLett.133.210401}%
  \BibitemOpen
  \bibfield  {author} {\bibinfo {author} {\bibfnamefont {T.}~\bibnamefont {Roscilde}}, \bibinfo {author} {\bibfnamefont {F.}~\bibnamefont {Caleca}}, \bibinfo {author} {\bibfnamefont {A.}~\bibnamefont {Angelone}},\ and\ \bibinfo {author} {\bibfnamefont {F.}~\bibnamefont {Mezzacapo}},\ }\bibfield  {title} {\bibinfo {title} {Scalable spin squeezing from critical slowing down in short-range interacting systems},\ }\href {https://doi.org/10.1103/PhysRevLett.133.210401} {\bibfield  {journal} {\bibinfo  {journal} {Phys. Rev. Lett.}\ }\textbf {\bibinfo {volume} {133}},\ \bibinfo {pages} {210401} (\bibinfo {year} {2024})}\BibitemShut {NoStop}%
\bibitem [{\citenamefont {Chaudhry}\ and\ \citenamefont {Gong}(2012)}]{PhysRevA.86.012311}%
  \BibitemOpen
  \bibfield  {author} {\bibinfo {author} {\bibfnamefont {A.~Z.}\ \bibnamefont {Chaudhry}}\ and\ \bibinfo {author} {\bibfnamefont {J.}~\bibnamefont {Gong}},\ }\bibfield  {title} {\bibinfo {title} {Protecting and enhancing spin squeezing via continuous dynamical decoupling},\ }\href {https://doi.org/10.1103/PhysRevA.86.012311} {\bibfield  {journal} {\bibinfo  {journal} {Phys. Rev. A}\ }\textbf {\bibinfo {volume} {86}},\ \bibinfo {pages} {012311} (\bibinfo {year} {2012})}\BibitemShut {NoStop}%
\bibitem [{\citenamefont {Liao}\ \emph {et~al.}(2017)\citenamefont {Liao}, \citenamefont {Rong},\ and\ \citenamefont {Fang}}]{Liao_2017}%
  \BibitemOpen
  \bibfield  {author} {\bibinfo {author} {\bibfnamefont {X.-P.}\ \bibnamefont {Liao}}, \bibinfo {author} {\bibfnamefont {M.-S.}\ \bibnamefont {Rong}},\ and\ \bibinfo {author} {\bibfnamefont {M.-F.}\ \bibnamefont {Fang}},\ }\bibfield  {title} {\bibinfo {title} {Protecting and enhancing spin squeezing from decoherence using weak measurements},\ }\href {https://doi.org/10.1088/1612-202X/aa6dc7} {\bibfield  {journal} {\bibinfo  {journal} {Laser Physics Letters}\ }\textbf {\bibinfo {volume} {14}},\ \bibinfo {pages} {065201} (\bibinfo {year} {2017})}\BibitemShut {NoStop}%
\bibitem [{\citenamefont {Bai}\ and\ \citenamefont {An}(2021)}]{PhysRevLett.127.083602}%
  \BibitemOpen
  \bibfield  {author} {\bibinfo {author} {\bibfnamefont {S.-Y.}\ \bibnamefont {Bai}}\ and\ \bibinfo {author} {\bibfnamefont {J.-H.}\ \bibnamefont {An}},\ }\bibfield  {title} {\bibinfo {title} {Generating stable spin squeezing by squeezed-reservoir engineering},\ }\href {https://doi.org/10.1103/PhysRevLett.127.083602} {\bibfield  {journal} {\bibinfo  {journal} {Phys. Rev. Lett.}\ }\textbf {\bibinfo {volume} {127}},\ \bibinfo {pages} {083602} (\bibinfo {year} {2021})}\BibitemShut {NoStop}%
\bibitem [{\citenamefont {Matsuzaki}\ \emph {et~al.}(2011)\citenamefont {Matsuzaki}, \citenamefont {Benjamin},\ and\ \citenamefont {Fitzsimons}}]{PhysRevA.84.012103}%
  \BibitemOpen
  \bibfield  {author} {\bibinfo {author} {\bibfnamefont {Y.}~\bibnamefont {Matsuzaki}}, \bibinfo {author} {\bibfnamefont {S.~C.}\ \bibnamefont {Benjamin}},\ and\ \bibinfo {author} {\bibfnamefont {J.}~\bibnamefont {Fitzsimons}},\ }\bibfield  {title} {\bibinfo {title} {Magnetic field sensing beyond the standard quantum limit under the effect of decoherence},\ }\href {https://doi.org/10.1103/PhysRevA.84.012103} {\bibfield  {journal} {\bibinfo  {journal} {Phys. Rev. A}\ }\textbf {\bibinfo {volume} {84}},\ \bibinfo {pages} {012103} (\bibinfo {year} {2011})}\BibitemShut {NoStop}%
\bibitem [{\citenamefont {Tanaka}\ \emph {et~al.}(2015)\citenamefont {Tanaka}, \citenamefont {Knott}, \citenamefont {Matsuzaki}, \citenamefont {Dooley}, \citenamefont {Yamaguchi}, \citenamefont {Munro},\ and\ \citenamefont {Saito}}]{PhysRevLett.115.170801}%
  \BibitemOpen
  \bibfield  {author} {\bibinfo {author} {\bibfnamefont {T.}~\bibnamefont {Tanaka}}, \bibinfo {author} {\bibfnamefont {P.}~\bibnamefont {Knott}}, \bibinfo {author} {\bibfnamefont {Y.}~\bibnamefont {Matsuzaki}}, \bibinfo {author} {\bibfnamefont {S.}~\bibnamefont {Dooley}}, \bibinfo {author} {\bibfnamefont {H.}~\bibnamefont {Yamaguchi}}, \bibinfo {author} {\bibfnamefont {W.~J.}\ \bibnamefont {Munro}},\ and\ \bibinfo {author} {\bibfnamefont {S.}~\bibnamefont {Saito}},\ }\bibfield  {title} {\bibinfo {title} {Proposed robust entanglement-based magnetic field sensor beyond the standard quantum limit},\ }\href {https://doi.org/10.1103/PhysRevLett.115.170801} {\bibfield  {journal} {\bibinfo  {journal} {Phys. Rev. Lett.}\ }\textbf {\bibinfo {volume} {115}},\ \bibinfo {pages} {170801} (\bibinfo {year} {2015})}\BibitemShut {NoStop}%
\bibitem [{\citenamefont {Rivas}\ \emph {et~al.}(2014)\citenamefont {Rivas}, \citenamefont {Huelga},\ and\ \citenamefont {Plenio}}]{Rivas_2014}%
  \BibitemOpen
  \bibfield  {author} {\bibinfo {author} {\bibfnamefont {A.}~\bibnamefont {Rivas}}, \bibinfo {author} {\bibfnamefont {S.~F.}\ \bibnamefont {Huelga}},\ and\ \bibinfo {author} {\bibfnamefont {M.~B.}\ \bibnamefont {Plenio}},\ }\bibfield  {title} {\bibinfo {title} {Quantum non-{M}arkovianity: characterization, quantification and detection},\ }\href {https://doi.org/10.1088/0034-4885/77/9/094001} {\bibfield  {journal} {\bibinfo  {journal} {Reports on Progress in Physics}\ }\textbf {\bibinfo {volume} {77}},\ \bibinfo {pages} {094001} (\bibinfo {year} {2014})}\BibitemShut {NoStop}%
\bibitem [{\citenamefont {Breuer}\ \emph {et~al.}(2016)\citenamefont {Breuer}, \citenamefont {Laine}, \citenamefont {Piilo},\ and\ \citenamefont {Vacchini}}]{RevModPhys.88.021002}%
  \BibitemOpen
  \bibfield  {author} {\bibinfo {author} {\bibfnamefont {H.-P.}\ \bibnamefont {Breuer}}, \bibinfo {author} {\bibfnamefont {E.-M.}\ \bibnamefont {Laine}}, \bibinfo {author} {\bibfnamefont {J.}~\bibnamefont {Piilo}},\ and\ \bibinfo {author} {\bibfnamefont {B.}~\bibnamefont {Vacchini}},\ }\bibfield  {title} {\bibinfo {title} {Colloquium: Non-{M}arkovian dynamics in open quantum systems},\ }\href {https://doi.org/10.1103/RevModPhys.88.021002} {\bibfield  {journal} {\bibinfo  {journal} {Rev. Mod. Phys.}\ }\textbf {\bibinfo {volume} {88}},\ \bibinfo {pages} {021002} (\bibinfo {year} {2016})}\BibitemShut {NoStop}%
\bibitem [{\citenamefont {de~Vega}\ and\ \citenamefont {Alonso}(2017)}]{RevModPhys.89.015001}%
  \BibitemOpen
  \bibfield  {author} {\bibinfo {author} {\bibfnamefont {I.}~\bibnamefont {de~Vega}}\ and\ \bibinfo {author} {\bibfnamefont {D.}~\bibnamefont {Alonso}},\ }\bibfield  {title} {\bibinfo {title} {Dynamics of non-{M}arkovian open quantum systems},\ }\href {https://doi.org/10.1103/RevModPhys.89.015001} {\bibfield  {journal} {\bibinfo  {journal} {Rev. Mod. Phys.}\ }\textbf {\bibinfo {volume} {89}},\ \bibinfo {pages} {015001} (\bibinfo {year} {2017})}\BibitemShut {NoStop}%
\bibitem [{\citenamefont {Li}\ \emph {et~al.}(2018{\natexlab{a}})\citenamefont {Li}, \citenamefont {Hall},\ and\ \citenamefont {Wiseman}}]{LI20181}%
  \BibitemOpen
  \bibfield  {author} {\bibinfo {author} {\bibfnamefont {L.}~\bibnamefont {Li}}, \bibinfo {author} {\bibfnamefont {M.~J.}\ \bibnamefont {Hall}},\ and\ \bibinfo {author} {\bibfnamefont {H.~M.}\ \bibnamefont {Wiseman}},\ }\bibfield  {title} {\bibinfo {title} {Concepts of quantum non-{M}arkovianity: A hierarchy},\ }\href {https://doi.org/https://doi.org/10.1016/j.physrep.2018.07.001} {\bibfield  {journal} {\bibinfo  {journal} {Physics Reports}\ }\textbf {\bibinfo {volume} {759}},\ \bibinfo {pages} {1} (\bibinfo {year} {2018}{\natexlab{a}})}\BibitemShut {NoStop}%
\bibitem [{\citenamefont {Bai}\ \emph {et~al.}(2019)\citenamefont {Bai}, \citenamefont {Peng}, \citenamefont {Luo},\ and\ \citenamefont {An}}]{PhysRevLett.123.040402}%
  \BibitemOpen
  \bibfield  {author} {\bibinfo {author} {\bibfnamefont {K.}~\bibnamefont {Bai}}, \bibinfo {author} {\bibfnamefont {Z.}~\bibnamefont {Peng}}, \bibinfo {author} {\bibfnamefont {H.-G.}\ \bibnamefont {Luo}},\ and\ \bibinfo {author} {\bibfnamefont {J.-H.}\ \bibnamefont {An}},\ }\bibfield  {title} {\bibinfo {title} {Retrieving ideal precision in noisy quantum optical metrology},\ }\href {https://doi.org/10.1103/PhysRevLett.123.040402} {\bibfield  {journal} {\bibinfo  {journal} {Phys. Rev. Lett.}\ }\textbf {\bibinfo {volume} {123}},\ \bibinfo {pages} {040402} (\bibinfo {year} {2019})}\BibitemShut {NoStop}%
\bibitem [{\citenamefont {Chin}\ \emph {et~al.}(2012)\citenamefont {Chin}, \citenamefont {Huelga},\ and\ \citenamefont {Plenio}}]{PhysRevLett.109.233601}%
  \BibitemOpen
  \bibfield  {author} {\bibinfo {author} {\bibfnamefont {A.~W.}\ \bibnamefont {Chin}}, \bibinfo {author} {\bibfnamefont {S.~F.}\ \bibnamefont {Huelga}},\ and\ \bibinfo {author} {\bibfnamefont {M.~B.}\ \bibnamefont {Plenio}},\ }\bibfield  {title} {\bibinfo {title} {Quantum metrology in non-{M}arkovian environments},\ }\href {https://doi.org/10.1103/PhysRevLett.109.233601} {\bibfield  {journal} {\bibinfo  {journal} {Phys. Rev. Lett.}\ }\textbf {\bibinfo {volume} {109}},\ \bibinfo {pages} {233601} (\bibinfo {year} {2012})}\BibitemShut {NoStop}%
\bibitem [{\citenamefont {Long}\ \emph {et~al.}(2022)\citenamefont {Long}, \citenamefont {He}, \citenamefont {Zhang}, \citenamefont {Tang}, \citenamefont {Lin}, \citenamefont {Liu}, \citenamefont {Nie}, \citenamefont {Feng}, \citenamefont {Li}, \citenamefont {Xin}, \citenamefont {Ai},\ and\ \citenamefont {Lu}}]{PhysRevLett.129.070502}%
  \BibitemOpen
  \bibfield  {author} {\bibinfo {author} {\bibfnamefont {X.}~\bibnamefont {Long}}, \bibinfo {author} {\bibfnamefont {W.-T.}\ \bibnamefont {He}}, \bibinfo {author} {\bibfnamefont {N.-N.}\ \bibnamefont {Zhang}}, \bibinfo {author} {\bibfnamefont {K.}~\bibnamefont {Tang}}, \bibinfo {author} {\bibfnamefont {Z.}~\bibnamefont {Lin}}, \bibinfo {author} {\bibfnamefont {H.}~\bibnamefont {Liu}}, \bibinfo {author} {\bibfnamefont {X.}~\bibnamefont {Nie}}, \bibinfo {author} {\bibfnamefont {G.}~\bibnamefont {Feng}}, \bibinfo {author} {\bibfnamefont {J.}~\bibnamefont {Li}}, \bibinfo {author} {\bibfnamefont {T.}~\bibnamefont {Xin}}, \bibinfo {author} {\bibfnamefont {Q.}~\bibnamefont {Ai}},\ and\ \bibinfo {author} {\bibfnamefont {D.}~\bibnamefont {Lu}},\ }\bibfield  {title} {\bibinfo {title} {Entanglement-enhanced quantum metrology in colored noise by quantum {Z}eno effect},\ }\href {https://doi.org/10.1103/PhysRevLett.129.070502} {\bibfield  {journal} {\bibinfo  {journal} {Phys. Rev. Lett.}\ }\textbf {\bibinfo {volume}
  {129}},\ \bibinfo {pages} {070502} (\bibinfo {year} {2022})}\BibitemShut {NoStop}%
\bibitem [{\citenamefont {Bai}\ and\ \citenamefont {An}(2023)}]{PhysRevLett.131.050801}%
  \BibitemOpen
  \bibfield  {author} {\bibinfo {author} {\bibfnamefont {S.-Y.}\ \bibnamefont {Bai}}\ and\ \bibinfo {author} {\bibfnamefont {J.-H.}\ \bibnamefont {An}},\ }\bibfield  {title} {\bibinfo {title} {Floquet engineering to overcome no-go theorem of noisy quantum metrology},\ }\href {https://doi.org/10.1103/PhysRevLett.131.050801} {\bibfield  {journal} {\bibinfo  {journal} {Phys. Rev. Lett.}\ }\textbf {\bibinfo {volume} {131}},\ \bibinfo {pages} {050801} (\bibinfo {year} {2023})}\BibitemShut {NoStop}%
\bibitem [{\citenamefont {Itano}\ \emph {et~al.}(1993)\citenamefont {Itano}, \citenamefont {Bergquist}, \citenamefont {Bollinger}, \citenamefont {Gilligan}, \citenamefont {Heinzen}, \citenamefont {Moore}, \citenamefont {Raizen},\ and\ \citenamefont {Wineland}}]{PhysRevA.47.3554}%
  \BibitemOpen
  \bibfield  {author} {\bibinfo {author} {\bibfnamefont {W.~M.}\ \bibnamefont {Itano}}, \bibinfo {author} {\bibfnamefont {J.~C.}\ \bibnamefont {Bergquist}}, \bibinfo {author} {\bibfnamefont {J.~J.}\ \bibnamefont {Bollinger}}, \bibinfo {author} {\bibfnamefont {J.~M.}\ \bibnamefont {Gilligan}}, \bibinfo {author} {\bibfnamefont {D.~J.}\ \bibnamefont {Heinzen}}, \bibinfo {author} {\bibfnamefont {F.~L.}\ \bibnamefont {Moore}}, \bibinfo {author} {\bibfnamefont {M.~G.}\ \bibnamefont {Raizen}},\ and\ \bibinfo {author} {\bibfnamefont {D.~J.}\ \bibnamefont {Wineland}},\ }\bibfield  {title} {\bibinfo {title} {Quantum projection noise: Population fluctuations in two-level systems},\ }\href {https://doi.org/10.1103/PhysRevA.47.3554} {\bibfield  {journal} {\bibinfo  {journal} {Phys. Rev. A}\ }\textbf {\bibinfo {volume} {47}},\ \bibinfo {pages} {3554} (\bibinfo {year} {1993})}\BibitemShut {NoStop}%
\bibitem [{\citenamefont {Sanders}(1989)}]{PhysRevA.40.2417}%
  \BibitemOpen
  \bibfield  {author} {\bibinfo {author} {\bibfnamefont {B.~C.}\ \bibnamefont {Sanders}},\ }\bibfield  {title} {\bibinfo {title} {Quantum dynamics of the nonlinear rotator and the effects of continual spin measurement},\ }\href {https://doi.org/10.1103/PhysRevA.40.2417} {\bibfield  {journal} {\bibinfo  {journal} {Phys. Rev. A}\ }\textbf {\bibinfo {volume} {40}},\ \bibinfo {pages} {2417} (\bibinfo {year} {1989})}\BibitemShut {NoStop}%
\bibitem [{\citenamefont {Wu}\ \emph {et~al.}(2015)\citenamefont {Wu}, \citenamefont {Jin},\ and\ \citenamefont {You}}]{PhysRevA.92.033826}%
  \BibitemOpen
  \bibfield  {author} {\bibinfo {author} {\bibfnamefont {L.-N.}\ \bibnamefont {Wu}}, \bibinfo {author} {\bibfnamefont {G.-R.}\ \bibnamefont {Jin}},\ and\ \bibinfo {author} {\bibfnamefont {L.}~\bibnamefont {You}},\ }\bibfield  {title} {\bibinfo {title} {Spin squeezing of the non-{H}ermitian one-axis twisting model},\ }\href {https://doi.org/10.1103/PhysRevA.92.033826} {\bibfield  {journal} {\bibinfo  {journal} {Phys. Rev. A}\ }\textbf {\bibinfo {volume} {92}},\ \bibinfo {pages} {033826} (\bibinfo {year} {2015})}\BibitemShut {NoStop}%
\bibitem [{\citenamefont {Fujiwara}(2001)}]{PhysRevA.63.042304}%
  \BibitemOpen
  \bibfield  {author} {\bibinfo {author} {\bibfnamefont {A.}~\bibnamefont {Fujiwara}},\ }\bibfield  {title} {\bibinfo {title} {Quantum channel identification problem},\ }\href {https://doi.org/10.1103/PhysRevA.63.042304} {\bibfield  {journal} {\bibinfo  {journal} {Phys. Rev. A}\ }\textbf {\bibinfo {volume} {63}},\ \bibinfo {pages} {042304} (\bibinfo {year} {2001})}\BibitemShut {NoStop}%
\bibitem [{\citenamefont {Fernholz}\ \emph {et~al.}(2008)\citenamefont {Fernholz}, \citenamefont {Krauter}, \citenamefont {Jensen}, \citenamefont {Sherson}, \citenamefont {S\o{}rensen},\ and\ \citenamefont {Polzik}}]{PhysRevLett.101.073601}%
  \BibitemOpen
  \bibfield  {author} {\bibinfo {author} {\bibfnamefont {T.}~\bibnamefont {Fernholz}}, \bibinfo {author} {\bibfnamefont {H.}~\bibnamefont {Krauter}}, \bibinfo {author} {\bibfnamefont {K.}~\bibnamefont {Jensen}}, \bibinfo {author} {\bibfnamefont {J.~F.}\ \bibnamefont {Sherson}}, \bibinfo {author} {\bibfnamefont {A.~S.}\ \bibnamefont {S\o{}rensen}},\ and\ \bibinfo {author} {\bibfnamefont {E.~S.}\ \bibnamefont {Polzik}},\ }\bibfield  {title} {\bibinfo {title} {Spin squeezing of atomic ensembles via nuclear-electronic spin entanglement},\ }\href {https://doi.org/10.1103/PhysRevLett.101.073601} {\bibfield  {journal} {\bibinfo  {journal} {Phys. Rev. Lett.}\ }\textbf {\bibinfo {volume} {101}},\ \bibinfo {pages} {073601} (\bibinfo {year} {2008})}\BibitemShut {NoStop}%
\bibitem [{\citenamefont {Chaudhury}\ \emph {et~al.}(2007)\citenamefont {Chaudhury}, \citenamefont {Merkel}, \citenamefont {Herr}, \citenamefont {Silberfarb}, \citenamefont {Deutsch},\ and\ \citenamefont {Jessen}}]{PhysRevLett.99.163002}%
  \BibitemOpen
  \bibfield  {author} {\bibinfo {author} {\bibfnamefont {S.}~\bibnamefont {Chaudhury}}, \bibinfo {author} {\bibfnamefont {S.}~\bibnamefont {Merkel}}, \bibinfo {author} {\bibfnamefont {T.}~\bibnamefont {Herr}}, \bibinfo {author} {\bibfnamefont {A.}~\bibnamefont {Silberfarb}}, \bibinfo {author} {\bibfnamefont {I.~H.}\ \bibnamefont {Deutsch}},\ and\ \bibinfo {author} {\bibfnamefont {P.~S.}\ \bibnamefont {Jessen}},\ }\bibfield  {title} {\bibinfo {title} {Quantum control of the hyperfine spin of a {C}s atom ensemble},\ }\href {https://doi.org/10.1103/PhysRevLett.99.163002} {\bibfield  {journal} {\bibinfo  {journal} {Phys. Rev. Lett.}\ }\textbf {\bibinfo {volume} {99}},\ \bibinfo {pages} {163002} (\bibinfo {year} {2007})}\BibitemShut {NoStop}%
\bibitem [{\citenamefont {Grond}\ \emph {et~al.}(2009)\citenamefont {Grond}, \citenamefont {Schmiedmayer},\ and\ \citenamefont {Hohenester}}]{PhysRevA.79.021603}%
  \BibitemOpen
  \bibfield  {author} {\bibinfo {author} {\bibfnamefont {J.}~\bibnamefont {Grond}}, \bibinfo {author} {\bibfnamefont {J.}~\bibnamefont {Schmiedmayer}},\ and\ \bibinfo {author} {\bibfnamefont {U.}~\bibnamefont {Hohenester}},\ }\bibfield  {title} {\bibinfo {title} {Optimizing number squeezing when splitting a mesoscopic condensate},\ }\href {https://doi.org/10.1103/PhysRevA.79.021603} {\bibfield  {journal} {\bibinfo  {journal} {Phys. Rev. A}\ }\textbf {\bibinfo {volume} {79}},\ \bibinfo {pages} {021603} (\bibinfo {year} {2009})}\BibitemShut {NoStop}%
\bibitem [{\citenamefont {Orzel}\ \emph {et~al.}(2001)\citenamefont {Orzel}, \citenamefont {Tuchman}, \citenamefont {Fenselau}, \citenamefont {Yasuda},\ and\ \citenamefont {Kasevich}}]{doi:10.1126/science.1058149}%
  \BibitemOpen
  \bibfield  {author} {\bibinfo {author} {\bibfnamefont {C.}~\bibnamefont {Orzel}}, \bibinfo {author} {\bibfnamefont {A.~K.}\ \bibnamefont {Tuchman}}, \bibinfo {author} {\bibfnamefont {M.~L.}\ \bibnamefont {Fenselau}}, \bibinfo {author} {\bibfnamefont {M.}~\bibnamefont {Yasuda}},\ and\ \bibinfo {author} {\bibfnamefont {M.~A.}\ \bibnamefont {Kasevich}},\ }\bibfield  {title} {\bibinfo {title} {Squeezed states in a {B}ose-{E}instein condensate},\ }\href {https://doi.org/10.1126/science.1058149} {\bibfield  {journal} {\bibinfo  {journal} {Science}\ }\textbf {\bibinfo {volume} {291}},\ \bibinfo {pages} {2386} (\bibinfo {year} {2001})}\BibitemShut {NoStop}%
\bibitem [{\citenamefont {Bohnet}\ \emph {et~al.}(2016)\citenamefont {Bohnet}, \citenamefont {Sawyer}, \citenamefont {Britton}, \citenamefont {Wall}, \citenamefont {Rey}, \citenamefont {Foss-Feig},\ and\ \citenamefont {Bollinger}}]{doi:10.1126/science.aad9958}%
  \BibitemOpen
  \bibfield  {author} {\bibinfo {author} {\bibfnamefont {J.~G.}\ \bibnamefont {Bohnet}}, \bibinfo {author} {\bibfnamefont {B.~C.}\ \bibnamefont {Sawyer}}, \bibinfo {author} {\bibfnamefont {J.~W.}\ \bibnamefont {Britton}}, \bibinfo {author} {\bibfnamefont {M.~L.}\ \bibnamefont {Wall}}, \bibinfo {author} {\bibfnamefont {A.~M.}\ \bibnamefont {Rey}}, \bibinfo {author} {\bibfnamefont {M.}~\bibnamefont {Foss-Feig}},\ and\ \bibinfo {author} {\bibfnamefont {J.~J.}\ \bibnamefont {Bollinger}},\ }\bibfield  {title} {\bibinfo {title} {Quantum spin dynamics and entanglement generation with hundreds of trapped ions},\ }\href {https://doi.org/10.1126/science.aad9958} {\bibfield  {journal} {\bibinfo  {journal} {Science}\ }\textbf {\bibinfo {volume} {352}},\ \bibinfo {pages} {1297} (\bibinfo {year} {2016})}\BibitemShut {NoStop}%
\bibitem [{\citenamefont {Lu}\ \emph {et~al.}(2019)\citenamefont {Lu}, \citenamefont {Zhang}, \citenamefont {Zhang}, \citenamefont {Chen}, \citenamefont {Shen}, \citenamefont {Zhang}, \citenamefont {Zhang},\ and\ \citenamefont {Kim}}]{Lu2019}%
  \BibitemOpen
  \bibfield  {author} {\bibinfo {author} {\bibfnamefont {Y.}~\bibnamefont {Lu}}, \bibinfo {author} {\bibfnamefont {S.}~\bibnamefont {Zhang}}, \bibinfo {author} {\bibfnamefont {K.}~\bibnamefont {Zhang}}, \bibinfo {author} {\bibfnamefont {W.}~\bibnamefont {Chen}}, \bibinfo {author} {\bibfnamefont {Y.}~\bibnamefont {Shen}}, \bibinfo {author} {\bibfnamefont {J.}~\bibnamefont {Zhang}}, \bibinfo {author} {\bibfnamefont {J.-N.}\ \bibnamefont {Zhang}},\ and\ \bibinfo {author} {\bibfnamefont {K.}~\bibnamefont {Kim}},\ }\bibfield  {title} {\bibinfo {title} {Global entangling gates on arbitrary ion qubits},\ }\href {https://doi.org/10.1038/s41586-019-1428-4} {\bibfield  {journal} {\bibinfo  {journal} {Nature}\ }\textbf {\bibinfo {volume} {572}},\ \bibinfo {pages} {363} (\bibinfo {year} {2019})}\BibitemShut {NoStop}%
\bibitem [{\citenamefont {Figgatt}\ \emph {et~al.}(2019)\citenamefont {Figgatt}, \citenamefont {Ostrander}, \citenamefont {Linke}, \citenamefont {Landsman}, \citenamefont {Zhu}, \citenamefont {Maslov},\ and\ \citenamefont {Monroe}}]{Figgatt2019}%
  \BibitemOpen
  \bibfield  {author} {\bibinfo {author} {\bibfnamefont {C.}~\bibnamefont {Figgatt}}, \bibinfo {author} {\bibfnamefont {A.}~\bibnamefont {Ostrander}}, \bibinfo {author} {\bibfnamefont {N.~M.}\ \bibnamefont {Linke}}, \bibinfo {author} {\bibfnamefont {K.~A.}\ \bibnamefont {Landsman}}, \bibinfo {author} {\bibfnamefont {D.}~\bibnamefont {Zhu}}, \bibinfo {author} {\bibfnamefont {D.}~\bibnamefont {Maslov}},\ and\ \bibinfo {author} {\bibfnamefont {C.}~\bibnamefont {Monroe}},\ }\bibfield  {title} {\bibinfo {title} {Parallel entangling operations on a universal ion-trap quantum computer},\ }\href {https://doi.org/10.1038/s41586-019-1427-5} {\bibfield  {journal} {\bibinfo  {journal} {Nature}\ }\textbf {\bibinfo {volume} {572}},\ \bibinfo {pages} {368} (\bibinfo {year} {2019})}\BibitemShut {NoStop}%
\bibitem [{\citenamefont {Song}\ \emph {et~al.}(2019)\citenamefont {Song}, \citenamefont {Xu}, \citenamefont {Li}, \citenamefont {Zhang}, \citenamefont {Zhang}, \citenamefont {Liu}, \citenamefont {Guo}, \citenamefont {Wang}, \citenamefont {Ren}, \citenamefont {Hao}, \citenamefont {Feng}, \citenamefont {Fan}, \citenamefont {Zheng}, \citenamefont {Wang}, \citenamefont {Wang},\ and\ \citenamefont {Zhu}}]{doi:10.1126/science.aay0600}%
  \BibitemOpen
  \bibfield  {author} {\bibinfo {author} {\bibfnamefont {C.}~\bibnamefont {Song}}, \bibinfo {author} {\bibfnamefont {K.}~\bibnamefont {Xu}}, \bibinfo {author} {\bibfnamefont {H.}~\bibnamefont {Li}}, \bibinfo {author} {\bibfnamefont {Y.-R.}\ \bibnamefont {Zhang}}, \bibinfo {author} {\bibfnamefont {X.}~\bibnamefont {Zhang}}, \bibinfo {author} {\bibfnamefont {W.}~\bibnamefont {Liu}}, \bibinfo {author} {\bibfnamefont {Q.}~\bibnamefont {Guo}}, \bibinfo {author} {\bibfnamefont {Z.}~\bibnamefont {Wang}}, \bibinfo {author} {\bibfnamefont {W.}~\bibnamefont {Ren}}, \bibinfo {author} {\bibfnamefont {J.}~\bibnamefont {Hao}}, \bibinfo {author} {\bibfnamefont {H.}~\bibnamefont {Feng}}, \bibinfo {author} {\bibfnamefont {H.}~\bibnamefont {Fan}}, \bibinfo {author} {\bibfnamefont {D.}~\bibnamefont {Zheng}}, \bibinfo {author} {\bibfnamefont {D.-W.}\ \bibnamefont {Wang}}, \bibinfo {author} {\bibfnamefont {H.}~\bibnamefont {Wang}},\ and\ \bibinfo {author} {\bibfnamefont {S.-Y.}\ \bibnamefont {Zhu}},\ }\bibfield  {title} {\bibinfo
  {title} {Generation of multicomponent atomic {S}chrödinger cat states of up to 20 qubits},\ }\href {https://doi.org/10.1126/science.aay0600} {\bibfield  {journal} {\bibinfo  {journal} {Science}\ }\textbf {\bibinfo {volume} {365}},\ \bibinfo {pages} {574} (\bibinfo {year} {2019})}\BibitemShut {NoStop}%
\bibitem [{\citenamefont {Xu}\ \emph {et~al.}(2020)\citenamefont {Xu}, \citenamefont {Sun}, \citenamefont {Liu}, \citenamefont {Zhang}, \citenamefont {Li}, \citenamefont {Dong}, \citenamefont {Ren}, \citenamefont {Zhang}, \citenamefont {Nori}, \citenamefont {Zheng}, \citenamefont {Fan},\ and\ \citenamefont {Wang}}]{doi:10.1126/sciadv.aba4935}%
  \BibitemOpen
  \bibfield  {author} {\bibinfo {author} {\bibfnamefont {K.}~\bibnamefont {Xu}}, \bibinfo {author} {\bibfnamefont {Z.-H.}\ \bibnamefont {Sun}}, \bibinfo {author} {\bibfnamefont {W.}~\bibnamefont {Liu}}, \bibinfo {author} {\bibfnamefont {Y.-R.}\ \bibnamefont {Zhang}}, \bibinfo {author} {\bibfnamefont {H.}~\bibnamefont {Li}}, \bibinfo {author} {\bibfnamefont {H.}~\bibnamefont {Dong}}, \bibinfo {author} {\bibfnamefont {W.}~\bibnamefont {Ren}}, \bibinfo {author} {\bibfnamefont {P.}~\bibnamefont {Zhang}}, \bibinfo {author} {\bibfnamefont {F.}~\bibnamefont {Nori}}, \bibinfo {author} {\bibfnamefont {D.}~\bibnamefont {Zheng}}, \bibinfo {author} {\bibfnamefont {H.}~\bibnamefont {Fan}},\ and\ \bibinfo {author} {\bibfnamefont {H.}~\bibnamefont {Wang}},\ }\bibfield  {title} {\bibinfo {title} {Probing dynamical phase transitions with a superconducting quantum simulator},\ }\href {https://doi.org/10.1126/sciadv.aba4935} {\bibfield  {journal} {\bibinfo  {journal} {Science Advances}\ }\textbf {\bibinfo {volume} {6}},\
  \bibinfo {pages} {eaba4935} (\bibinfo {year} {2020})}\BibitemShut {NoStop}%
\bibitem [{\citenamefont {Liu}\ \emph {et~al.}(2011)\citenamefont {Liu}, \citenamefont {Xu}, \citenamefont {Jin},\ and\ \citenamefont {You}}]{PhysRevLett.107.013601}%
  \BibitemOpen
  \bibfield  {author} {\bibinfo {author} {\bibfnamefont {Y.~C.}\ \bibnamefont {Liu}}, \bibinfo {author} {\bibfnamefont {Z.~F.}\ \bibnamefont {Xu}}, \bibinfo {author} {\bibfnamefont {G.~R.}\ \bibnamefont {Jin}},\ and\ \bibinfo {author} {\bibfnamefont {L.}~\bibnamefont {You}},\ }\bibfield  {title} {\bibinfo {title} {Spin squeezing: Transforming one-axis twisting into two-axis twisting},\ }\href {https://doi.org/10.1103/PhysRevLett.107.013601} {\bibfield  {journal} {\bibinfo  {journal} {Phys. Rev. Lett.}\ }\textbf {\bibinfo {volume} {107}},\ \bibinfo {pages} {013601} (\bibinfo {year} {2011})}\BibitemShut {NoStop}%
\bibitem [{\citenamefont {Zhang}\ \emph {et~al.}(2014)\citenamefont {Zhang}, \citenamefont {Zhou}, \citenamefont {Guo},\ and\ \citenamefont {Zhou}}]{PhysRevA.90.013604}%
  \BibitemOpen
  \bibfield  {author} {\bibinfo {author} {\bibfnamefont {J.-Y.}\ \bibnamefont {Zhang}}, \bibinfo {author} {\bibfnamefont {X.-F.}\ \bibnamefont {Zhou}}, \bibinfo {author} {\bibfnamefont {G.-C.}\ \bibnamefont {Guo}},\ and\ \bibinfo {author} {\bibfnamefont {Z.-W.}\ \bibnamefont {Zhou}},\ }\bibfield  {title} {\bibinfo {title} {Dynamical spin squeezing via a higher-order {T}rotter-{S}uzuki approximation},\ }\href {https://doi.org/10.1103/PhysRevA.90.013604} {\bibfield  {journal} {\bibinfo  {journal} {Phys. Rev. A}\ }\textbf {\bibinfo {volume} {90}},\ \bibinfo {pages} {013604} (\bibinfo {year} {2014})}\BibitemShut {NoStop}%
\bibitem [{\citenamefont {Huang}\ \emph {et~al.}(2015)\citenamefont {Huang}, \citenamefont {Zhang}, \citenamefont {Zou}, \citenamefont {Zou},\ and\ \citenamefont {Guo}}]{PhysRevA.91.043642}%
  \BibitemOpen
  \bibfield  {author} {\bibinfo {author} {\bibfnamefont {W.}~\bibnamefont {Huang}}, \bibinfo {author} {\bibfnamefont {Y.-L.}\ \bibnamefont {Zhang}}, \bibinfo {author} {\bibfnamefont {C.-L.}\ \bibnamefont {Zou}}, \bibinfo {author} {\bibfnamefont {X.-B.}\ \bibnamefont {Zou}},\ and\ \bibinfo {author} {\bibfnamefont {G.-C.}\ \bibnamefont {Guo}},\ }\bibfield  {title} {\bibinfo {title} {Two-axis spin squeezing of two-component {B}ose-{E}instein condensates via continuous driving},\ }\href {https://doi.org/10.1103/PhysRevA.91.043642} {\bibfield  {journal} {\bibinfo  {journal} {Phys. Rev. A}\ }\textbf {\bibinfo {volume} {91}},\ \bibinfo {pages} {043642} (\bibinfo {year} {2015})}\BibitemShut {NoStop}%
\bibitem [{\citenamefont {Huang}\ \emph {et~al.}(2023{\natexlab{b}})\citenamefont {Huang}, \citenamefont {Zhang}, \citenamefont {Wang}, \citenamefont {Hua}, \citenamefont {Tang},\ and\ \citenamefont {Liu}}]{PhysRevA.107.042613}%
  \BibitemOpen
  \bibfield  {author} {\bibinfo {author} {\bibfnamefont {L.-G.}\ \bibnamefont {Huang}}, \bibinfo {author} {\bibfnamefont {X.}~\bibnamefont {Zhang}}, \bibinfo {author} {\bibfnamefont {Y.}~\bibnamefont {Wang}}, \bibinfo {author} {\bibfnamefont {Z.}~\bibnamefont {Hua}}, \bibinfo {author} {\bibfnamefont {Y.}~\bibnamefont {Tang}},\ and\ \bibinfo {author} {\bibfnamefont {Y.-C.}\ \bibnamefont {Liu}},\ }\bibfield  {title} {\bibinfo {title} {Heisenberg-limited spin squeezing in coupled spin systems},\ }\href {https://doi.org/10.1103/PhysRevA.107.042613} {\bibfield  {journal} {\bibinfo  {journal} {Phys. Rev. A}\ }\textbf {\bibinfo {volume} {107}},\ \bibinfo {pages} {042613} (\bibinfo {year} {2023}{\natexlab{b}})}\BibitemShut {NoStop}%
\bibitem [{\citenamefont {Xu}\ \emph {et~al.}(2017)\citenamefont {Xu}, \citenamefont {Sun}, \citenamefont {Yi},\ and\ \citenamefont {Zhang}}]{Xu2017}%
  \BibitemOpen
  \bibfield  {author} {\bibinfo {author} {\bibfnamefont {P.}~\bibnamefont {Xu}}, \bibinfo {author} {\bibfnamefont {H.}~\bibnamefont {Sun}}, \bibinfo {author} {\bibfnamefont {S.}~\bibnamefont {Yi}},\ and\ \bibinfo {author} {\bibfnamefont {W.}~\bibnamefont {Zhang}},\ }\bibfield  {title} {\bibinfo {title} {Rebuilding of destroyed spin squeezing in noisy environments},\ }\href {https://doi.org/10.1038/s41598-017-14442-5} {\bibfield  {journal} {\bibinfo  {journal} {Sci. Rep.}\ }\textbf {\bibinfo {volume} {7}},\ \bibinfo {pages} {14102} (\bibinfo {year} {2017})}\BibitemShut {NoStop}%
\bibitem [{\citenamefont {Luo}\ \emph {et~al.}(2025)\citenamefont {Luo}, \citenamefont {Zhang}, \citenamefont {Chu}, \citenamefont {Maruko}, \citenamefont {Rey},\ and\ \citenamefont {Thompson}}]{Luo2025}%
  \BibitemOpen
  \bibfield  {author} {\bibinfo {author} {\bibfnamefont {C.}~\bibnamefont {Luo}}, \bibinfo {author} {\bibfnamefont {H.}~\bibnamefont {Zhang}}, \bibinfo {author} {\bibfnamefont {A.}~\bibnamefont {Chu}}, \bibinfo {author} {\bibfnamefont {C.}~\bibnamefont {Maruko}}, \bibinfo {author} {\bibfnamefont {A.~M.}\ \bibnamefont {Rey}},\ and\ \bibinfo {author} {\bibfnamefont {J.~K.}\ \bibnamefont {Thompson}},\ }\bibfield  {title} {\bibinfo {title} {Hamiltonian engineering of collective xyz spin models in an optical cavity},\ }\href {https://doi.org/10.1038/s41567-025-02866-0} {\bibfield  {journal} {\bibinfo  {journal} {Nature Physics}\ }\textbf {\bibinfo {volume} {21}},\ \bibinfo {pages} {916} (\bibinfo {year} {2025})}\BibitemShut {NoStop}%
\bibitem [{\citenamefont {Tong}\ \emph {et~al.}(2010)\citenamefont {Tong}, \citenamefont {An}, \citenamefont {Luo},\ and\ \citenamefont {Oh}}]{PhysRevA.81.052330}%
  \BibitemOpen
  \bibfield  {author} {\bibinfo {author} {\bibfnamefont {Q.-J.}\ \bibnamefont {Tong}}, \bibinfo {author} {\bibfnamefont {J.-H.}\ \bibnamefont {An}}, \bibinfo {author} {\bibfnamefont {H.-G.}\ \bibnamefont {Luo}},\ and\ \bibinfo {author} {\bibfnamefont {C.~H.}\ \bibnamefont {Oh}},\ }\bibfield  {title} {\bibinfo {title} {Mechanism of entanglement preservation},\ }\href {https://doi.org/10.1103/PhysRevA.81.052330} {\bibfield  {journal} {\bibinfo  {journal} {Phys. Rev. A}\ }\textbf {\bibinfo {volume} {81}},\ \bibinfo {pages} {052330} (\bibinfo {year} {2010})}\BibitemShut {NoStop}%
\bibitem [{\citenamefont {Zhang}\ \emph {et~al.}(2012)\citenamefont {Zhang}, \citenamefont {Lo}, \citenamefont {Xiong}, \citenamefont {Tu},\ and\ \citenamefont {Nori}}]{PhysRevLett.109.170402}%
  \BibitemOpen
  \bibfield  {author} {\bibinfo {author} {\bibfnamefont {W.-M.}\ \bibnamefont {Zhang}}, \bibinfo {author} {\bibfnamefont {P.-Y.}\ \bibnamefont {Lo}}, \bibinfo {author} {\bibfnamefont {H.-N.}\ \bibnamefont {Xiong}}, \bibinfo {author} {\bibfnamefont {M.~W.-Y.}\ \bibnamefont {Tu}},\ and\ \bibinfo {author} {\bibfnamefont {F.}~\bibnamefont {Nori}},\ }\bibfield  {title} {\bibinfo {title} {General non-{M}arkovian dynamics of open quantum systems},\ }\href {https://doi.org/10.1103/PhysRevLett.109.170402} {\bibfield  {journal} {\bibinfo  {journal} {Phys. Rev. Lett.}\ }\textbf {\bibinfo {volume} {109}},\ \bibinfo {pages} {170402} (\bibinfo {year} {2012})}\BibitemShut {NoStop}%
\bibitem [{\citenamefont {Zhu}\ \emph {et~al.}(2018)\citenamefont {Zhu}, \citenamefont {Zhang}, \citenamefont {Zhuang},\ and\ \citenamefont {Liu}}]{PhysRevLett.121.220403}%
  \BibitemOpen
  \bibfield  {author} {\bibinfo {author} {\bibfnamefont {H.-J.}\ \bibnamefont {Zhu}}, \bibinfo {author} {\bibfnamefont {G.-F.}\ \bibnamefont {Zhang}}, \bibinfo {author} {\bibfnamefont {L.}~\bibnamefont {Zhuang}},\ and\ \bibinfo {author} {\bibfnamefont {W.-M.}\ \bibnamefont {Liu}},\ }\bibfield  {title} {\bibinfo {title} {Universal dissipationless dynamics in {G}aussian continuous-variable open systems},\ }\href {https://doi.org/10.1103/PhysRevLett.121.220403} {\bibfield  {journal} {\bibinfo  {journal} {Phys. Rev. Lett.}\ }\textbf {\bibinfo {volume} {121}},\ \bibinfo {pages} {220403} (\bibinfo {year} {2018})}\BibitemShut {NoStop}%
\bibitem [{\citenamefont {Scharnagl}\ \emph {et~al.}(2023)\citenamefont {Scharnagl}, \citenamefont {Kielinski},\ and\ \citenamefont {Hammerer}}]{PhysRevA.108.062611}%
  \BibitemOpen
  \bibfield  {author} {\bibinfo {author} {\bibfnamefont {M.~S.}\ \bibnamefont {Scharnagl}}, \bibinfo {author} {\bibfnamefont {T.}~\bibnamefont {Kielinski}},\ and\ \bibinfo {author} {\bibfnamefont {K.}~\bibnamefont {Hammerer}},\ }\bibfield  {title} {\bibinfo {title} {Optimal ramsey interferometry with echo protocols based on one-axis twisting},\ }\href {https://doi.org/10.1103/PhysRevA.108.062611} {\bibfield  {journal} {\bibinfo  {journal} {Phys. Rev. A}\ }\textbf {\bibinfo {volume} {108}},\ \bibinfo {pages} {062611} (\bibinfo {year} {2023})}\BibitemShut {NoStop}%
\bibitem [{\citenamefont {Schulte}\ \emph {et~al.}(2020{\natexlab{b}})\citenamefont {Schulte}, \citenamefont {Mart{\'{i}}nez-Lahuerta}, \citenamefont {Scharnagl},\ and\ \citenamefont {Hammerer}}]{Schulte2020ramsey}%
  \BibitemOpen
  \bibfield  {author} {\bibinfo {author} {\bibfnamefont {M.}~\bibnamefont {Schulte}}, \bibinfo {author} {\bibfnamefont {V.~J.}\ \bibnamefont {Mart{\'{i}}nez-Lahuerta}}, \bibinfo {author} {\bibfnamefont {M.~S.}\ \bibnamefont {Scharnagl}},\ and\ \bibinfo {author} {\bibfnamefont {K.}~\bibnamefont {Hammerer}},\ }\bibfield  {title} {\bibinfo {title} {Ramsey interferometry with generalized one-axis twisting echoes},\ }\href {https://doi.org/10.22331/q-2020-05-15-268} {\bibfield  {journal} {\bibinfo  {journal} {{Quantum}}\ }\textbf {\bibinfo {volume} {4}},\ \bibinfo {pages} {268} (\bibinfo {year} {2020}{\natexlab{b}})}\BibitemShut {NoStop}%
\bibitem [{\citenamefont {Sun}\ \emph {et~al.}(2025)\citenamefont {Sun}, \citenamefont {Kang}, \citenamefont {Nuomin}, \citenamefont {Schwartz}, \citenamefont {Beratan}, \citenamefont {Brown},\ and\ \citenamefont {Kim}}]{Sun2025}%
  \BibitemOpen
  \bibfield  {author} {\bibinfo {author} {\bibfnamefont {K.}~\bibnamefont {Sun}}, \bibinfo {author} {\bibfnamefont {M.}~\bibnamefont {Kang}}, \bibinfo {author} {\bibfnamefont {H.}~\bibnamefont {Nuomin}}, \bibinfo {author} {\bibfnamefont {G.}~\bibnamefont {Schwartz}}, \bibinfo {author} {\bibfnamefont {D.~N.}\ \bibnamefont {Beratan}}, \bibinfo {author} {\bibfnamefont {K.~R.}\ \bibnamefont {Brown}},\ and\ \bibinfo {author} {\bibfnamefont {J.}~\bibnamefont {Kim}},\ }\bibfield  {title} {\bibinfo {title} {Quantum simulation of spin-boson models with structured bath},\ }\href {https://doi.org/10.1038/s41467-025-59296-y} {\bibfield  {journal} {\bibinfo  {journal} {Nature Communications}\ }\textbf {\bibinfo {volume} {16}},\ \bibinfo {pages} {4042} (\bibinfo {year} {2025})}\BibitemShut {NoStop}%
\bibitem [{\citenamefont {Forn-Díaz}\ \emph {et~al.}(2017)\citenamefont {Forn-Díaz}, \citenamefont {García-Ripoll}, \citenamefont {Peropadre}, \citenamefont {Orgiazzi}, \citenamefont {Yurtalan}, \citenamefont {Belyansky}, \citenamefont {Wilson},\ and\ \citenamefont {Lupascu}}]{FornDiaz2017}%
  \BibitemOpen
  \bibfield  {author} {\bibinfo {author} {\bibfnamefont {P.}~\bibnamefont {Forn-Díaz}}, \bibinfo {author} {\bibfnamefont {J.~J.}\ \bibnamefont {García-Ripoll}}, \bibinfo {author} {\bibfnamefont {B.}~\bibnamefont {Peropadre}}, \bibinfo {author} {\bibfnamefont {J.-L.}\ \bibnamefont {Orgiazzi}}, \bibinfo {author} {\bibfnamefont {M.~A.}\ \bibnamefont {Yurtalan}}, \bibinfo {author} {\bibfnamefont {R.}~\bibnamefont {Belyansky}}, \bibinfo {author} {\bibfnamefont {C.~M.}\ \bibnamefont {Wilson}},\ and\ \bibinfo {author} {\bibfnamefont {A.}~\bibnamefont {Lupascu}},\ }\bibfield  {title} {\bibinfo {title} {Ultrastrong coupling of a single artificial atom to an electromagnetic continuum in the nonperturbative regime},\ }\href {https://doi.org/10.1038/nphys3905} {\bibfield  {journal} {\bibinfo  {journal} {Nature Physics}\ }\textbf {\bibinfo {volume} {13}},\ \bibinfo {pages} {39} (\bibinfo {year} {2017})}\BibitemShut {NoStop}%
\bibitem [{\citenamefont {Magazzù}\ \emph {et~al.}(2018)\citenamefont {Magazzù}, \citenamefont {Forn-Díaz}, \citenamefont {Belyansky}, \citenamefont {Orgiazzi}, \citenamefont {Yurtalan}, \citenamefont {Otto}, \citenamefont {Lupascu}, \citenamefont {Wilson},\ and\ \citenamefont {Grifoni}}]{Magazzù2018}%
  \BibitemOpen
  \bibfield  {author} {\bibinfo {author} {\bibfnamefont {L.}~\bibnamefont {Magazzù}}, \bibinfo {author} {\bibfnamefont {P.}~\bibnamefont {Forn-Díaz}}, \bibinfo {author} {\bibfnamefont {R.}~\bibnamefont {Belyansky}}, \bibinfo {author} {\bibfnamefont {J.-L.}\ \bibnamefont {Orgiazzi}}, \bibinfo {author} {\bibfnamefont {M.~A.}\ \bibnamefont {Yurtalan}}, \bibinfo {author} {\bibfnamefont {M.~R.}\ \bibnamefont {Otto}}, \bibinfo {author} {\bibfnamefont {A.}~\bibnamefont {Lupascu}}, \bibinfo {author} {\bibfnamefont {C.~M.}\ \bibnamefont {Wilson}},\ and\ \bibinfo {author} {\bibfnamefont {M.}~\bibnamefont {Grifoni}},\ }\bibfield  {title} {\bibinfo {title} {Probing the strongly driven spin-boson model in a superconducting quantum circuit},\ }\href {https://doi.org/10.1038/s41467-018-03626-w} {\bibfield  {journal} {\bibinfo  {journal} {Nature Communications}\ }\textbf {\bibinfo {volume} {9}},\ \bibinfo {pages} {1403} (\bibinfo {year} {2018})}\BibitemShut {NoStop}%
\bibitem [{\citenamefont {Leggett}\ \emph {et~al.}(1987)\citenamefont {Leggett}, \citenamefont {Chakravarty}, \citenamefont {Dorsey}, \citenamefont {Fisher}, \citenamefont {Garg},\ and\ \citenamefont {Zwerger}}]{RevModPhys.59.1}%
  \BibitemOpen
  \bibfield  {author} {\bibinfo {author} {\bibfnamefont {A.~J.}\ \bibnamefont {Leggett}}, \bibinfo {author} {\bibfnamefont {S.}~\bibnamefont {Chakravarty}}, \bibinfo {author} {\bibfnamefont {A.~T.}\ \bibnamefont {Dorsey}}, \bibinfo {author} {\bibfnamefont {M.~P.~A.}\ \bibnamefont {Fisher}}, \bibinfo {author} {\bibfnamefont {A.}~\bibnamefont {Garg}},\ and\ \bibinfo {author} {\bibfnamefont {W.}~\bibnamefont {Zwerger}},\ }\bibfield  {title} {\bibinfo {title} {Dynamics of the dissipative two-state system},\ }\href {https://doi.org/10.1103/RevModPhys.59.1} {\bibfield  {journal} {\bibinfo  {journal} {Rev. Mod. Phys.}\ }\textbf {\bibinfo {volume} {59}},\ \bibinfo {pages} {1} (\bibinfo {year} {1987})}\BibitemShut {NoStop}%
\bibitem [{\citenamefont {Kraus}\ \emph {et~al.}(1983)\citenamefont {Kraus}, \citenamefont {Böhm}, \citenamefont {Dollard},\ and\ \citenamefont {Wootters}}]{Kraus1983StatesEA}%
  \BibitemOpen
  \bibfield  {author} {\bibinfo {author} {\bibfnamefont {K.}~\bibnamefont {Kraus}}, \bibinfo {author} {\bibfnamefont {A.}~\bibnamefont {Böhm}}, \bibinfo {author} {\bibfnamefont {J.~D.}\ \bibnamefont {Dollard}},\ and\ \bibinfo {author} {\bibfnamefont {W.~H.}\ \bibnamefont {Wootters}},\ }\href {https://doi.org/10.1007/3540127321_22} {\emph {\bibinfo {title} {States, Effects, and Operations Fundamental Notions of Quantum Theory: Lectures in Mathematical Physics at the University of Texas at Austin}}}\ (\bibinfo  {publisher} {Springer Berlin Heidelberg},\ \bibinfo {address} {Berlin, Heidelberg},\ \bibinfo {year} {1983})\BibitemShut {NoStop}%
\bibitem [{\citenamefont {Dorner}\ \emph {et~al.}(2009)\citenamefont {Dorner}, \citenamefont {Demkowicz-Dobrzanski}, \citenamefont {Smith}, \citenamefont {Lundeen}, \citenamefont {Wasilewski}, \citenamefont {Banaszek},\ and\ \citenamefont {Walmsley}}]{PhysRevLett.102.040403}%
  \BibitemOpen
  \bibfield  {author} {\bibinfo {author} {\bibfnamefont {U.}~\bibnamefont {Dorner}}, \bibinfo {author} {\bibfnamefont {R.}~\bibnamefont {Demkowicz-Dobrzanski}}, \bibinfo {author} {\bibfnamefont {B.~J.}\ \bibnamefont {Smith}}, \bibinfo {author} {\bibfnamefont {J.~S.}\ \bibnamefont {Lundeen}}, \bibinfo {author} {\bibfnamefont {W.}~\bibnamefont {Wasilewski}}, \bibinfo {author} {\bibfnamefont {K.}~\bibnamefont {Banaszek}},\ and\ \bibinfo {author} {\bibfnamefont {I.~A.}\ \bibnamefont {Walmsley}},\ }\bibfield  {title} {\bibinfo {title} {Optimal quantum phase estimation},\ }\href {https://doi.org/10.1103/PhysRevLett.102.040403} {\bibfield  {journal} {\bibinfo  {journal} {Phys. Rev. Lett.}\ }\textbf {\bibinfo {volume} {102}},\ \bibinfo {pages} {040403} (\bibinfo {year} {2009})}\BibitemShut {NoStop}%
\bibitem [{\citenamefont {Cooper}\ \emph {et~al.}(2012)\citenamefont {Cooper}, \citenamefont {Hallwood}, \citenamefont {Dunningham},\ and\ \citenamefont {Brand}}]{PhysRevLett.108.130402}%
  \BibitemOpen
  \bibfield  {author} {\bibinfo {author} {\bibfnamefont {J.~J.}\ \bibnamefont {Cooper}}, \bibinfo {author} {\bibfnamefont {D.~W.}\ \bibnamefont {Hallwood}}, \bibinfo {author} {\bibfnamefont {J.~A.}\ \bibnamefont {Dunningham}},\ and\ \bibinfo {author} {\bibfnamefont {J.}~\bibnamefont {Brand}},\ }\bibfield  {title} {\bibinfo {title} {Robust quantum enhanced phase estimation in a multimode interferometer},\ }\href {https://doi.org/10.1103/PhysRevLett.108.130402} {\bibfield  {journal} {\bibinfo  {journal} {Phys. Rev. Lett.}\ }\textbf {\bibinfo {volume} {108}},\ \bibinfo {pages} {130402} (\bibinfo {year} {2012})}\BibitemShut {NoStop}%
\bibitem [{\citenamefont {Huang}\ \emph {et~al.}(2017)\citenamefont {Huang}, \citenamefont {Motes}, \citenamefont {Anisimov}, \citenamefont {Dowling},\ and\ \citenamefont {Berry}}]{PhysRevA.95.053837}%
  \BibitemOpen
  \bibfield  {author} {\bibinfo {author} {\bibfnamefont {Z.}~\bibnamefont {Huang}}, \bibinfo {author} {\bibfnamefont {K.~R.}\ \bibnamefont {Motes}}, \bibinfo {author} {\bibfnamefont {P.~M.}\ \bibnamefont {Anisimov}}, \bibinfo {author} {\bibfnamefont {J.~P.}\ \bibnamefont {Dowling}},\ and\ \bibinfo {author} {\bibfnamefont {D.~W.}\ \bibnamefont {Berry}},\ }\bibfield  {title} {\bibinfo {title} {Adaptive phase estimation with two-mode squeezed vacuum and parity measurement},\ }\href {https://doi.org/10.1103/PhysRevA.95.053837} {\bibfield  {journal} {\bibinfo  {journal} {Phys. Rev. A}\ }\textbf {\bibinfo {volume} {95}},\ \bibinfo {pages} {053837} (\bibinfo {year} {2017})}\BibitemShut {NoStop}%
\bibitem [{\citenamefont {Knott}\ \emph {et~al.}(2014)\citenamefont {Knott}, \citenamefont {Proctor}, \citenamefont {Nemoto}, \citenamefont {Dunningham},\ and\ \citenamefont {Munro}}]{PhysRevA.90.033846}%
  \BibitemOpen
  \bibfield  {author} {\bibinfo {author} {\bibfnamefont {P.~A.}\ \bibnamefont {Knott}}, \bibinfo {author} {\bibfnamefont {T.~J.}\ \bibnamefont {Proctor}}, \bibinfo {author} {\bibfnamefont {K.}~\bibnamefont {Nemoto}}, \bibinfo {author} {\bibfnamefont {J.~A.}\ \bibnamefont {Dunningham}},\ and\ \bibinfo {author} {\bibfnamefont {W.~J.}\ \bibnamefont {Munro}},\ }\bibfield  {title} {\bibinfo {title} {Effect of multimode entanglement on lossy optical quantum metrology},\ }\href {https://doi.org/10.1103/PhysRevA.90.033846} {\bibfield  {journal} {\bibinfo  {journal} {Phys. Rev. A}\ }\textbf {\bibinfo {volume} {90}},\ \bibinfo {pages} {033846} (\bibinfo {year} {2014})}\BibitemShut {NoStop}%
\bibitem [{\citenamefont {Wu}\ and\ \citenamefont {An}(2024)}]{PhysRevLett.133.050401}%
  \BibitemOpen
  \bibfield  {author} {\bibinfo {author} {\bibfnamefont {W.}~\bibnamefont {Wu}}\ and\ \bibinfo {author} {\bibfnamefont {J.-H.}\ \bibnamefont {An}},\ }\bibfield  {title} {\bibinfo {title} {Generalized quantum fluctuation theorem for energy exchange},\ }\href {https://doi.org/10.1103/PhysRevLett.133.050401} {\bibfield  {journal} {\bibinfo  {journal} {Phys. Rev. Lett.}\ }\textbf {\bibinfo {volume} {133}},\ \bibinfo {pages} {050401} (\bibinfo {year} {2024})}\BibitemShut {NoStop}%
\bibitem [{\citenamefont {Liu}\ and\ \citenamefont {Houck}(2012)}]{nat.phys.1}%
  \BibitemOpen
  \bibfield  {author} {\bibinfo {author} {\bibfnamefont {Y.}~\bibnamefont {Liu}}\ and\ \bibinfo {author} {\bibfnamefont {A.~A.}\ \bibnamefont {Houck}},\ }\bibfield  {title} {\bibinfo {title} {Quantum electrodynamics near a photonic bandgap},\ }\href {https://doi.org/10.1038/nphys3834} {\bibfield  {journal} {\bibinfo  {journal} {Nat. Phys.}\ }\textbf {\bibinfo {volume} {13}},\ \bibinfo {pages} {48} (\bibinfo {year} {2012})}\BibitemShut {NoStop}%
\bibitem [{\citenamefont {Krinner}\ \emph {et~al.}(2018)\citenamefont {Krinner}, \citenamefont {Stewart}, \citenamefont {Pazmiño}, \citenamefont {Kwon},\ and\ \citenamefont {Schneble}}]{nature5}%
  \BibitemOpen
  \bibfield  {author} {\bibinfo {author} {\bibfnamefont {L.}~\bibnamefont {Krinner}}, \bibinfo {author} {\bibfnamefont {M.}~\bibnamefont {Stewart}}, \bibinfo {author} {\bibfnamefont {A.}~\bibnamefont {Pazmiño}}, \bibinfo {author} {\bibfnamefont {J.}~\bibnamefont {Kwon}},\ and\ \bibinfo {author} {\bibfnamefont {D.}~\bibnamefont {Schneble}},\ }\bibfield  {title} {\bibinfo {title} {Spontaneous emission of matter waves from a tunable open quantum system},\ }\href {https://doi.org/10.1038/s41586-018-0348-z} {\bibfield  {journal} {\bibinfo  {journal} {Nature}\ }\textbf {\bibinfo {volume} {559}},\ \bibinfo {pages} {589} (\bibinfo {year} {2018})}\BibitemShut {NoStop}%
\bibitem [{\citenamefont {Kwon}\ \emph {et~al.}(2022)\citenamefont {Kwon}, \citenamefont {Kim}, \citenamefont {Lanuza},\ and\ \citenamefont {Schneble}}]{Kwon2022}%
  \BibitemOpen
  \bibfield  {author} {\bibinfo {author} {\bibfnamefont {J.}~\bibnamefont {Kwon}}, \bibinfo {author} {\bibfnamefont {Y.}~\bibnamefont {Kim}}, \bibinfo {author} {\bibfnamefont {A.}~\bibnamefont {Lanuza}},\ and\ \bibinfo {author} {\bibfnamefont {D.}~\bibnamefont {Schneble}},\ }\bibfield  {title} {\bibinfo {title} {Formation of matter-wave polaritons in an optical lattice},\ }\href {https://doi.org/10.1038/s41567-022-01565-4} {\bibfield  {journal} {\bibinfo  {journal} {Nature Physics}\ }\textbf {\bibinfo {volume} {18}},\ \bibinfo {pages} {657} (\bibinfo {year} {2022})}\BibitemShut {NoStop}%
\bibitem [{\citenamefont {Kim}\ \emph {et~al.}(2025)\citenamefont {Kim}, \citenamefont {Lanuza},\ and\ \citenamefont {Schneble}}]{Kim2025}%
  \BibitemOpen
  \bibfield  {author} {\bibinfo {author} {\bibfnamefont {Y.}~\bibnamefont {Kim}}, \bibinfo {author} {\bibfnamefont {A.}~\bibnamefont {Lanuza}},\ and\ \bibinfo {author} {\bibfnamefont {D.}~\bibnamefont {Schneble}},\ }\bibfield  {title} {\bibinfo {title} {Super- and subradiant dynamics of quantum emitters mediated by atomic matter waves},\ }\href {https://doi.org/10.1038/s41567-024-02676-w} {\bibfield  {journal} {\bibinfo  {journal} {Nature Physics}\ }\textbf {\bibinfo {volume} {21}},\ \bibinfo {pages} {70} (\bibinfo {year} {2025})}\BibitemShut {NoStop}%
\bibitem [{\citenamefont {Hosten}\ \emph {et~al.}(2016)\citenamefont {Hosten}, \citenamefont {Engelsen}, \citenamefont {Krishnakumar},\ and\ \citenamefont {Kasevich}}]{Hosten2016}%
  \BibitemOpen
  \bibfield  {author} {\bibinfo {author} {\bibfnamefont {O.}~\bibnamefont {Hosten}}, \bibinfo {author} {\bibfnamefont {N.~J.}\ \bibnamefont {Engelsen}}, \bibinfo {author} {\bibfnamefont {R.}~\bibnamefont {Krishnakumar}},\ and\ \bibinfo {author} {\bibfnamefont {M.~A.}\ \bibnamefont {Kasevich}},\ }\bibfield  {title} {\bibinfo {title} {Measurement noise 100 times lower than the quantum-projection limit using entangled atoms},\ }\href {https://doi.org/10.1038/nature16176} {\bibfield  {journal} {\bibinfo  {journal} {Nature}\ }\textbf {\bibinfo {volume} {529}},\ \bibinfo {pages} {505} (\bibinfo {year} {2016})}\BibitemShut {NoStop}%
\bibitem [{\citenamefont {Tamascelli}\ \emph {et~al.}(2018)\citenamefont {Tamascelli}, \citenamefont {Smirne}, \citenamefont {Huelga},\ and\ \citenamefont {Plenio}}]{PhysRevLett.120.030402}%
  \BibitemOpen
  \bibfield  {author} {\bibinfo {author} {\bibfnamefont {D.}~\bibnamefont {Tamascelli}}, \bibinfo {author} {\bibfnamefont {A.}~\bibnamefont {Smirne}}, \bibinfo {author} {\bibfnamefont {S.~F.}\ \bibnamefont {Huelga}},\ and\ \bibinfo {author} {\bibfnamefont {M.~B.}\ \bibnamefont {Plenio}},\ }\bibfield  {title} {\bibinfo {title} {Nonperturbative treatment of non-{M}arkovian dynamics of open quantum systems},\ }\href {https://doi.org/10.1103/PhysRevLett.120.030402} {\bibfield  {journal} {\bibinfo  {journal} {Phys. Rev. Lett.}\ }\textbf {\bibinfo {volume} {120}},\ \bibinfo {pages} {030402} (\bibinfo {year} {2018})}\BibitemShut {NoStop}%
\bibitem [{\citenamefont {Lemmer}\ \emph {et~al.}(2018)\citenamefont {Lemmer}, \citenamefont {Cormick}, \citenamefont {Tamascelli}, \citenamefont {Schaetz}, \citenamefont {Huelga},\ and\ \citenamefont {Plenio}}]{Lemmer2018}%
  \BibitemOpen
  \bibfield  {author} {\bibinfo {author} {\bibfnamefont {A.}~\bibnamefont {Lemmer}}, \bibinfo {author} {\bibfnamefont {C.}~\bibnamefont {Cormick}}, \bibinfo {author} {\bibfnamefont {D.}~\bibnamefont {Tamascelli}}, \bibinfo {author} {\bibfnamefont {T.}~\bibnamefont {Schaetz}}, \bibinfo {author} {\bibfnamefont {S.~F.}\ \bibnamefont {Huelga}},\ and\ \bibinfo {author} {\bibfnamefont {M.~B.}\ \bibnamefont {Plenio}},\ }\bibfield  {title} {\bibinfo {title} {A trapped-ion simulator for spin-boson models with structured environments},\ }\href {https://doi.org/10.1088/1367-2630/aac87d} {\bibfield  {journal} {\bibinfo  {journal} {New J. Phys.}\ }\textbf {\bibinfo {volume} {20}},\ \bibinfo {pages} {073002} (\bibinfo {year} {2018})}\BibitemShut {NoStop}%
\bibitem [{\citenamefont {Franke}\ \emph {et~al.}(2023)\citenamefont {Franke}, \citenamefont {Muleady}, \citenamefont {Kaubruegger}, \citenamefont {Kranzl}, \citenamefont {Blatt}, \citenamefont {Rey}, \citenamefont {Joshi},\ and\ \citenamefont {Roos}}]{Franke2023}%
  \BibitemOpen
  \bibfield  {author} {\bibinfo {author} {\bibfnamefont {J.}~\bibnamefont {Franke}}, \bibinfo {author} {\bibfnamefont {S.~R.}\ \bibnamefont {Muleady}}, \bibinfo {author} {\bibfnamefont {R.}~\bibnamefont {Kaubruegger}}, \bibinfo {author} {\bibfnamefont {F.}~\bibnamefont {Kranzl}}, \bibinfo {author} {\bibfnamefont {R.}~\bibnamefont {Blatt}}, \bibinfo {author} {\bibfnamefont {A.~M.}\ \bibnamefont {Rey}}, \bibinfo {author} {\bibfnamefont {M.~K.}\ \bibnamefont {Joshi}},\ and\ \bibinfo {author} {\bibfnamefont {C.~F.}\ \bibnamefont {Roos}},\ }\bibfield  {title} {\bibinfo {title} {Quantum-enhanced sensing on optical transitions through finite-range interactions},\ }\href {https://doi.org/10.1038/s41586-023-06472-z} {\bibfield  {journal} {\bibinfo  {journal} {Nature}\ }\textbf {\bibinfo {volume} {621}},\ \bibinfo {pages} {740} (\bibinfo {year} {2023})}\BibitemShut {NoStop}%
\bibitem [{\citenamefont {Wang}\ \emph {et~al.}(2024)\citenamefont {Wang}, \citenamefont {Wu}, \citenamefont {Yao}, \citenamefont {Lian}, \citenamefont {Cheng}, \citenamefont {Xu}, \citenamefont {Zhang}, \citenamefont {Jiang}, \citenamefont {Xu}, \citenamefont {Qi}, \citenamefont {Hou}, \citenamefont {Zhou}, \citenamefont {He},\ and\ \citenamefont {Duan}}]{PhysRevA.109.062402}%
  \BibitemOpen
  \bibfield  {author} {\bibinfo {author} {\bibfnamefont {G.-X.}\ \bibnamefont {Wang}}, \bibinfo {author} {\bibfnamefont {Y.-K.}\ \bibnamefont {Wu}}, \bibinfo {author} {\bibfnamefont {R.}~\bibnamefont {Yao}}, \bibinfo {author} {\bibfnamefont {W.-Q.}\ \bibnamefont {Lian}}, \bibinfo {author} {\bibfnamefont {Z.-J.}\ \bibnamefont {Cheng}}, \bibinfo {author} {\bibfnamefont {Y.-L.}\ \bibnamefont {Xu}}, \bibinfo {author} {\bibfnamefont {C.}~\bibnamefont {Zhang}}, \bibinfo {author} {\bibfnamefont {Y.}~\bibnamefont {Jiang}}, \bibinfo {author} {\bibfnamefont {Y.-Z.}\ \bibnamefont {Xu}}, \bibinfo {author} {\bibfnamefont {B.-X.}\ \bibnamefont {Qi}}, \bibinfo {author} {\bibfnamefont {P.-Y.}\ \bibnamefont {Hou}}, \bibinfo {author} {\bibfnamefont {Z.-C.}\ \bibnamefont {Zhou}}, \bibinfo {author} {\bibfnamefont {L.}~\bibnamefont {He}},\ and\ \bibinfo {author} {\bibfnamefont {L.-M.}\ \bibnamefont {Duan}},\ }\bibfield  {title} {\bibinfo {title} {Simulating the spin-boson model with a controllable reservoir in an ion trap},\
  }\href {https://doi.org/10.1103/PhysRevA.109.062402} {\bibfield  {journal} {\bibinfo  {journal} {Phys. Rev. A}\ }\textbf {\bibinfo {volume} {109}},\ \bibinfo {pages} {062402} (\bibinfo {year} {2024})}\BibitemShut {NoStop}%
\bibitem [{\citenamefont {Bruzewicz}\ \emph {et~al.}(2016)\citenamefont {Bruzewicz}, \citenamefont {McConnell}, \citenamefont {Chiaverini},\ and\ \citenamefont {Sage}}]{Bruzewicz2016}%
  \BibitemOpen
  \bibfield  {author} {\bibinfo {author} {\bibfnamefont {C.~D.}\ \bibnamefont {Bruzewicz}}, \bibinfo {author} {\bibfnamefont {R.}~\bibnamefont {McConnell}}, \bibinfo {author} {\bibfnamefont {J.}~\bibnamefont {Chiaverini}},\ and\ \bibinfo {author} {\bibfnamefont {J.~M.}\ \bibnamefont {Sage}},\ }\bibfield  {title} {\bibinfo {title} {Scalable loading of a two-dimensional trapped-ion array},\ }\href {https://doi.org/10.1038/ncomms13005} {\bibfield  {journal} {\bibinfo  {journal} {Nature Communications}\ }\textbf {\bibinfo {volume} {7}},\ \bibinfo {pages} {13005} (\bibinfo {year} {2016})}\BibitemShut {NoStop}%
\bibitem [{\citenamefont {Xue}(2013)}]{XUE20131328}%
  \BibitemOpen
  \bibfield  {author} {\bibinfo {author} {\bibfnamefont {P.}~\bibnamefont {Xue}},\ }\bibfield  {title} {\bibinfo {title} {Non-{M}arkovian dynamics of spin squeezing},\ }\href {https://doi.org/https://doi.org/10.1016/j.physleta.2013.04.006} {\bibfield  {journal} {\bibinfo  {journal} {Physics Letters A}\ }\textbf {\bibinfo {volume} {377}},\ \bibinfo {pages} {1328} (\bibinfo {year} {2013})}\BibitemShut {NoStop}%
\bibitem [{\citenamefont {Li}\ \emph {et~al.}(2018{\natexlab{b}})\citenamefont {Li}, \citenamefont {Du},\ and\ \citenamefont {Liang}}]{Li_2018}%
  \BibitemOpen
  \bibfield  {author} {\bibinfo {author} {\bibfnamefont {J.-Q.}\ \bibnamefont {Li}}, \bibinfo {author} {\bibfnamefont {L.}~\bibnamefont {Du}},\ and\ \bibinfo {author} {\bibfnamefont {J.-Q.}\ \bibnamefont {Liang}},\ }\bibfield  {title} {\bibinfo {title} {Spin squeezing and pairwise entanglement under non-{M}arkovian environments with dynamical decoupling pulses},\ }\href {https://doi.org/10.1088/1555-6611/aac7e5} {\bibfield  {journal} {\bibinfo  {journal} {Laser Physics}\ }\textbf {\bibinfo {volume} {28}},\ \bibinfo {pages} {095202} (\bibinfo {year} {2018}{\natexlab{b}})}\BibitemShut {NoStop}%
\bibitem [{\citenamefont {Yin}\ \emph {et~al.}(2012)\citenamefont {Yin}, \citenamefont {Ma}, \citenamefont {Wang},\ and\ \citenamefont {Nori}}]{PhysRevA.86.012308}%
  \BibitemOpen
  \bibfield  {author} {\bibinfo {author} {\bibfnamefont {X.}~\bibnamefont {Yin}}, \bibinfo {author} {\bibfnamefont {J.}~\bibnamefont {Ma}}, \bibinfo {author} {\bibfnamefont {X.}~\bibnamefont {Wang}},\ and\ \bibinfo {author} {\bibfnamefont {F.}~\bibnamefont {Nori}},\ }\bibfield  {title} {\bibinfo {title} {Spin squeezing under non-{M}arkovian channels by the hierarchy equation method},\ }\href {https://doi.org/10.1103/PhysRevA.86.012308} {\bibfield  {journal} {\bibinfo  {journal} {Phys. Rev. A}\ }\textbf {\bibinfo {volume} {86}},\ \bibinfo {pages} {012308} (\bibinfo {year} {2012})}\BibitemShut {NoStop}%
\bibitem [{\citenamefont {Ma}\ \emph {et~al.}(2020)\citenamefont {Ma}, \citenamefont {Ding},\ and\ \citenamefont {Yu}}]{PhysRevA.101.022327}%
  \BibitemOpen
  \bibfield  {author} {\bibinfo {author} {\bibfnamefont {Y.-H.}\ \bibnamefont {Ma}}, \bibinfo {author} {\bibfnamefont {Q.-Z.}\ \bibnamefont {Ding}},\ and\ \bibinfo {author} {\bibfnamefont {T.}~\bibnamefont {Yu}},\ }\bibfield  {title} {\bibinfo {title} {Persistent spin squeezing of a dissipative one-axis twisting model embedded in a general thermal environment},\ }\href {https://doi.org/10.1103/PhysRevA.101.022327} {\bibfield  {journal} {\bibinfo  {journal} {Phys. Rev. A}\ }\textbf {\bibinfo {volume} {101}},\ \bibinfo {pages} {022327} (\bibinfo {year} {2020})}\BibitemShut {NoStop}%
\end{thebibliography}%
\end{document}